\documentclass[eprint,amsmath,amssymb,aps]{revtex4-2}
\usepackage{mathtools,bm}
\usepackage{graphicx}
\usepackage{xcolor}
\usepackage{hyperref}
\hypersetup{colorlinks=True,
citecolor=red,
linkcolor=blue, 
urlcolor=black}
\usepackage{changes}

\usepackage{algorithm}
\usepackage{algpseudocode}

\begin{document}

\preprint{APS/123-QED}

\title{Sedimentation of flexible fibers through an ordered array of pillars}

\author{Ursy Makanga}
\email{Contact author: umakanga@flatironinstitute.org}
\thanks{Present address: Center for Computational Biology, Flatiron Institute, Simons Foundation, New York, New York 10010, USA.}
\author{Camille Duprat}
\author{Blaise Delmotte}

\affiliation{LadHyX, CNRS, Ecole Polytechnique, Institut Polytechnique de Paris, 91128, Palaiseau, France.}

\date{\today}

\begin{abstract}

In this study, we extend our previous work [Makanga \textit{et al.}, Physical Review Fluids 8, 044303 (2023)] to investigate the effects of a periodic array of pillars on the settling dynamics of a flexible fiber in a viscous fluid. By means of numerical simulations, we show that the fiber trajectory falls at long times into one of two migration modes, ``zigzag" and ``displacement", which are determined by the short-time scattering dynamics induced by fiber-pillar interactions. In the zigzag mode, there is no average displacement relative to the direction of gravity; the center of mass of the fiber follows a quasi-straight line. Conversely, in the displacement mode, there is a net displacement relative to the direction of gravity; the fiber center of mass follows a cyclical skew-bumping path. We show how this long-time behavior can be predicted using a theoretical model with minimal ingredients. These findings, together with the long-time transport properties of the fiber, provide insights into future experiments, as well as the design of gravity-based deterministic lateral displacement (DLD) devices to sort elongated particles based on their geometric and/or mechanical properties.  

\end{abstract}

\maketitle

\section{Introduction}

The motion of fibers in a fluid is ubiquitous in various situations. Examples include the process of making pulp in paper industries \cite{lundellFluidMechanicsPapermaking2011a} and the transport of microplastic fibers through fluidic environments \cite{browneAccumulationMicroplasticShorelines2011a, engdahlSimulatingMobilityMicroplastics2018, reSheddingLightInvisible2019, kaneDispersionAccumulationUltimate2019, sutherlandFluidDynamicsChallenges2023}. Over the past decade, several studies have focused on fluid-fiber interactions in obstacle-free environments, yielding numerous conclusions \cite{duroureDynamicsFlexibleFibers2019}. However, there are many instances in nature and engineering applications where elongated particles must navigate a fluidic environment containing obstacles whose size is comparable to the particle characteristic length. For instance, bacteria often live in biofilms, which act as an antibiotic resistance, and their migration to secondary sites results in the formation of biofilm streamers -- large elastic filaments -- that can lead to clogging while transported through complex environments \cite{drescherBiofilmStreamersCause2013a}. Understanding bacterial migration is essential for biomedical applications to prevent infections. On the other hand, in engineering applications -- especially for particle sorting purposes -- fiber-obstacle interactions are relevant for the design of \textit{deterministic lateral displacement} (DLD) devices, as the obstacles affect the long-time transport properties of elongated particles (e.g., long, linear polymer chains such as DNA molecules) through repetitive collisions \cite{chouSortingDiffusionAsymmetric1999, kulrattanarakAnalysisMixedMotion2011, kulrattanarakMixedMotionDeterministic2011, salafiReviewDeterministicLateral2019, makanga2025device}. In all the aforementioned situations, the dynamics of the fiber result from the complex interplay between internal elastic stresses, contact forces, and hydrodynamic interactions with the surrounding obstacles. 

Recently, Chakrabarti \textit{et al.}\ \cite{chakrabartiTrappingGlidingVaulting2020}  and Li \textit{et al.}\ \cite{liMicrofluidicBandpassFilter2026} analyzed the dynamics of semi-flexible fibers, convected by a background flow, through a structured array of pillars. The two-dimensional numerical simulations of Chakrabarti \textit{et al.} demonstrated the potential for the chromatographic sorting of Brownian fibers by length \cite{chakrabartiTrappingGlidingVaulting2020}. The experiments and three-dimensional numerical simulations of Li \textit{et al.}\ \cite{liMicrofluidicBandpassFilter2026}, building on their earlier work on fiber scattering by a single pillar \cite{liDynamicsRigidFibers2024}, revealed a ``jump" mode for a specific range of fiber lengths, inducing lateral displacement across the pillar array, while shorter and longer fibers follow the flow direction through the lattice. They demonstrated that this size selectivity stems from a complex interplay between fiber deformation, flow-induced tension, and interactions with obstacles.

When settling under gravity in an obstacle-free viscous fluid, due to its drag anisotropy, a rigid fiber drifts laterally at a constant velocity and maintains its initial orientation with respect to the direction of gravity \cite{batchelorSlenderbodyTheoryParticles1970}. When the fiber is allowed to bend, owing to the inhomogeneous drag distribution along its length, the fiber reorients to reach an equilibrium configuration after having traveled a finite distance \cite{liSedimentationFlexibleFilaments2013a, marchettiDeformationFlexibleFiber2018}. The presence of an obstacle is therefore expected to significantly affect these dynamics. In our recent study \cite{makangaObstacleinducedLateralDispersion2023}, we showed that, while settling against a rigid obstacle of arbitrary shape, a flexible fiber can either glide or remain trapped on it. The trapping states lead to nontrivial conformations of the fiber that result from the balance between elastic stresses, gravity, and friction on the obstacle surface. On the other hand, gliding cases lead to a lateral displacement, whose magnitude is large (typically several fiber lengths) and depends on the mechanical and geometric properties of the fiber. 

In an array of rigid obstacles, it is clear that repetitive collisions with the obstacles, as well as hydrodynamic interactions, will affect the dynamics described above and, consequently, the trajectory of the fiber. To further explore the potential of pillar arrays for sorting fibers under gravity, we investigate the dynamics of a single flexible fiber settling through a periodic array of pillars in a quiescent viscous fluid. 

The paper is organized as follows. We describe the problem in Sec. \ref{sec: problem_description}. The dynamical equations governing the motion of the fiber through the array are derived and numerically solved in Sec. \ref{sec: model_and_numerical_methods}. We then perform a parametric study and discuss the results in Sec. \ref{sec: results_and_discussion}. Conclusions are drawn in Sec. \ref{sec: conclusions}.

\section{Problem description}
\label{sec: problem_description}
\begin{figure}[h]
   \centering
    \includegraphics[width=\textwidth]{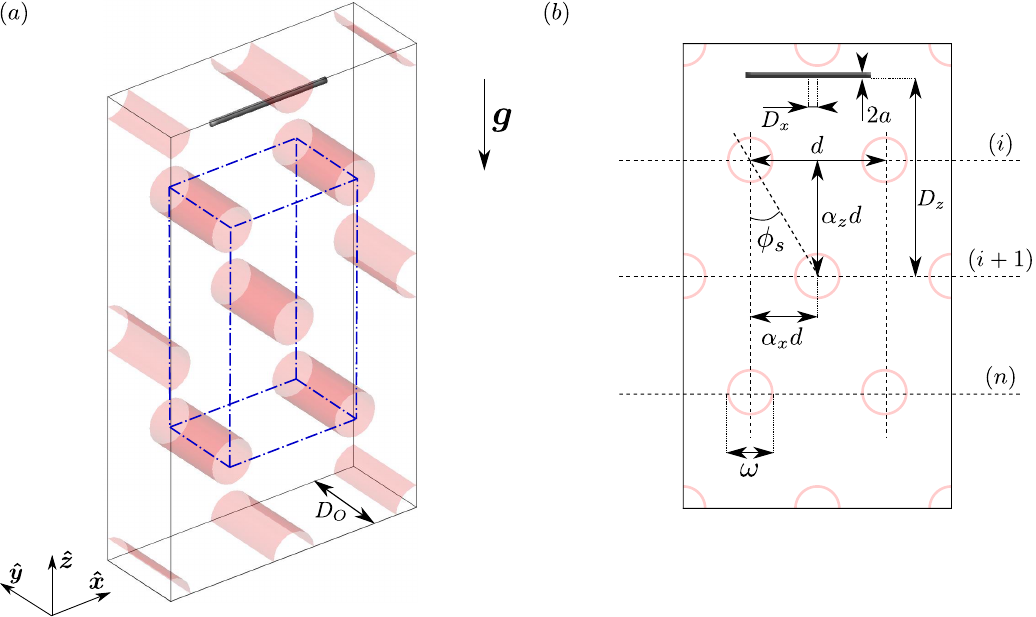}
    \caption{Schematic of the problem considered: A flexible fiber, initially straight, sediments under gravity in an array of rigid pillars, which is arranged in a hexagonal lattice, and immersed in a quiescent viscous fluid. $(a)$ A three-dimensional representation of the computational domain made of $2 \times 2$ unit cells. A unit cell (blue cuboid) contains the equivalent of two pillars of depth $D_O$. $(b)$ A planar representation of the computational domain that shows an array of pillars whose geometry is characterized by the parameters $\alpha_x=1/2$ and $\alpha_z=\sqrt{3}/2$.} 
    \label{fig: problemDescriptionSettlingInAnArrayOfPillars}
\end{figure}

We study the dynamics of a single flexible fiber settling under gravity through a periodic array of pillars, which is immersed in a quiescent viscous fluid of viscosity $\eta$. The fiber is an elastic rod of length $L$ with a circular cross section of radius $a$. The fiber is settling at a characteristic velocity $U$, such that the Reynolds number $\text{Re}=\rho UL/\eta \ll 1$, where $\rho$ is the mass density of the fluid. Therefore, inertial effects are negligible compared to viscous ones. The array is made of rigid cylindrical pillars of depth $D_O$ and width $\omega$, see Fig.\ \ref{fig: problemDescriptionSettlingInAnArrayOfPillars}$(a)$. The geometry of the array is generated following the commonly-used convention for the design of \textit{deterministic lateral displacement} (DLD) devices, used to sort particles in biological and microfluidic applications \cite{longMultidirectionalSortingModes2008, kimBrokenFlowSymmetry2017}. We denote by $d$ the center-to-center distance between two adjacent pillars, see Fig.\ \ref{fig: problemDescriptionSettlingInAnArrayOfPillars}$(b)$. Starting from row $(i)$, the subsequent row $(i+1)$ is placed downstream at a distance $\alpha_z d$ from the former, where $\alpha_z \in \mathbb{R}_{+}^{\ast}$. The pillars in this subsequent row are then shifted by a distance $\alpha_x d$ along the transverse direction, where $\alpha_x$ is known as the row-shift fraction. Following the same process, the $n$-th nearest row will be shifted by a distance $n\alpha_x d$ from row $(i)$, $n \in \mathbb{N}^{\ast}$. Thus, a periodic array is obtained for $\alpha_x=1/n$, and its structural angle $\phi_s$ is given by $\tan{(\phi_s)} = 1/(n\alpha_z)$. In this study, we set $\alpha_z=\sqrt{3}/2$ and $n=2$. This setting leads to an array of pillars arranged in a hexagonal lattice, which we characterize by the structural angle, $\phi_s = \pi/6$, and the lattice spacing relative to the width of the pillars, $d/\omega$. 

Following our previous study \cite{makangaObstacleinducedLateralDispersion2023}, in addition to the aforementioned geometric parameters, we introduce the relative length of the fiber with respect to the width of the pillars, $\xi=L/\omega$, as well as the elastogravitational number, $Be=F_G^{\mathrm{F}}L^2/EI$, where $E$ is the Young's modulus of the fiber, $I=\pi a^4/2$ the second moment of inertia of its cross section, and $F_G^{\mathrm{F}}=WL$ is the gravitational force with $W$ being the weight per unit length of the fiber. The elastogravitational number characterizes the deformation of the fiber at equilibrium when settling under gravity in a viscous fluid. It results from the balance between gravitational and elastic restoring forces. In an obstacle-free domain, when the value of $Be$ increases, the equilibrium shape of the fiber evolves from rod-like to U-like shapes \cite{marchettiDeformationFlexibleFiber2018}. 

\section{Model and numerical methods}
\label{sec: model_and_numerical_methods}

Simulations are performed using a numerical approach similar to that described in Ref.\ \cite{makangaObstacleinducedLateralDispersion2023}, which has been extensively validated against experiments. We summarize the approach here and refer the reader to Appendix \ref{app: details_on_the_numerical_method} for a more detailed description. The fiber is discretized as a series of evenly spaced $N_F$ markers located at $\bm{r}_i^{F} \in \mathbb{R}^{3}$, and connected by stiff linkers made of linear springs. In the absence of pillars, each marker, enclosed by a spherical bead of radius $a$, is subject to the following forces: (i) a gravitational force $\bm{F}^{F}_{G,i} = -2a W\hat{\bm{z}} \in \mathbb{R}^3$, and (ii) an internal elastic force $\bm{F}^{F}_{E, i} = -\bm{\nabla}_{\bm{r}^{F}_i} \mathcal{E} \in \mathbb{R}^3$, where the elastic potential $\mathcal{E}$ is given by

\begin{equation}
    \mathcal{E} = \underbrace{\frac{S}{4a}\sum_{i=2}^{N_F}\left(\lVert\bm{t}_i\rVert - 2a\right)^2}_{\mathcal{E}^S} + \underbrace{\frac{
    B}{2a}\sum_{i=2}^{N_F - 1} \left(1 - \hat{\bm{t}}_{i+1}\cdot\hat{\bm{t}}_{i}\right)}_{\mathcal{E}^B}
    \label{eq: elastic_potential_planar_deformations_settling_in_an_array},
\end{equation}
with $\bm{t}_i=\bm{r}^{F}_i - \bm{r}^{F}_{i-1}$ and  $\hat{\bm{t}}_i = \bm{t}_i/\lVert\bm{t}_i\rVert$ being respectively the tangential vector between two neighboring markers and the corresponding unit vector. In Eq.\ \eqref{eq: elastic_potential_planar_deformations_settling_in_an_array}, $S =\pi E a^2$ and $B= \pi E a^4/4$ correspond respectively to the stretching and bending moduli. Furthermore, the first term on the right-hand side of Eq.\ \eqref{eq: elastic_potential_planar_deformations_settling_in_an_array} is the stretching energy, $\mathcal{E}^S$, and the second term is the bending energy, $\mathcal{E}^B$.

On the other hand, each pillar of the array is discretized with $N_O$ rigidly connected markers located at $\bm{r}_i^{O}\in \mathbb{R}^3$ and enclosed by spherical beads of radius $a$. Each marker located on the pillars has zero velocity, since the pillars are immotile. This kinematic constraint is enforced by applying a constraint force $\bm{F}^{O}_{C,i} \in \mathbb{R}^3$ to each of these markers. Finally, owing to fiber-pillar collisions, a pairwise-repulsive force $\bm{F}_{R,i}^{\bullet} = \sum_j\bm{F}_{R,ij} \in \mathbb{R}^3$ is applied between the markers located on the fiber, $\bm{F}_{R,i}^{F}$, and those located on the pillars, $\bm{F}_{R,i}^{O}$, in order to prevent overlaps \cite{danceCollisionBarrierEffects2004, makangaObstacleinducedLateralDispersion2023}, where

\begin{equation}
    \bm{F}_{R,ij}=
   \begin{dcases}
     -\frac{\gamma_{R}}{2a}\left( \frac{R_{c}^{2} - r_{ij}^{2}}{R_{c}^{2} - 4a^{2}}\right)^{4}\bm{r}_{ij}, & \text{for } r_{ij} < R_{c}\\
      \bm{0}, & \text{otherwise}
\end{dcases}
   \label{eq: steric_def_forces}
\end{equation}
with $\bm{r}_{ij} = \bm{r}_{j}^{O} - \bm{r}_{i}^{F}$ and $r_{ij} = \lVert\bm{r}_{ij}\lVert$. In Eq. \eqref{eq: steric_def_forces}, $\gamma_{R} = 8aW$ and $R_{c} = 2.2a$ control the strength of repulsion and the cutoff distance over which the force acts, respectively.

In our model, a unit lattice cell contains the equivalent of two pillars (see Fig.\ \ref{fig: problemDescriptionSettlingInAnArrayOfPillars}$(a)$), while the computational domain consists of $2 \times 2$ unit cells, thus including the equivalent of $8$ pillars, plus the fiber. Therefore, the total number of markers in the computational domain is $N=N_F + 8N_O$. In the absence of a background flow, the kinematic equation -- which governs the dynamics of the fiber and pillars -- is given by the following mobility relation: 

\begin{equation}
 \frac{\mathrm{d}\bm{R}}{\mathrm{d}t} \equiv \bm{U} = \bm{\mathcal{M}} \cdot \left(\bm{F}_G + \bm{F}_E + \bm{F}_R + \bm{F}_C\right)
\label{eq: mob_problem_settlin_in_an_array},
\end{equation} 
where $\bm{R} = [\bm{r}_1^{F}, \ldots, \bm{r}_{N_F}^{F}, \bm{r}_{1}^{O}, \ldots,\bm{r}_{8N_O}^{O}]$, $\bm{U} = [\bm{U}_1^{F}, \ldots, \bm{U}_{N_F}^{F}, \bm{U}_{1}^{O}, \ldots,\bm{U}_{8N_O}^{O}]$, $\bm{F}_G = [\bm{F}_{G,1}^{F}, \ldots, \bm{F}_{G,N_F}^{F}, \bm{0}]$, $\bm{F}_E = [\bm{F}_{E,1}^{F}, \ldots, \bm{F}_{E,N_F}^{F}, \bm{0}]$, $\bm{F}_R = [\bm{F}_{R,1}^{F}, \ldots, \bm{F}_{R, N_F}^{F}, \bm{F}_{R, 1}^{O}, \ldots,\bm{F}_{R, 8N_O}^{O}]$, and $\bm{F}_C = [\bm{0}, \bm{F}_{C,1}^{O}, \ldots, \bm{F}_{C,8N_O}^{O}]$ are  $3N$ vectors collecting respectively the positions of the markers, their translational velocities, the gravitational, elastic, contact, and constraint forces. $\bm{\mathcal{M}}$ is the $3N\times3N$ mobility matrix that encodes hydrodynamic interactions between all the markers in the domain. This mobility matrix is given by the Rotne-Prager-Yamakawa (RPY) hydrodynamic tensor \cite{rotneVariationalTreatmentHydrodynamic1969, wajnrybGeneralizationRotnePragerYamakawaMobility2013, zukRotnePragerYamakawa2014} in a triply periodic domain \cite{mizerskiRotnePragerYamakawaApproximationPeriodic2014, fioreRapidSamplingStochastic2017}. Here, we use the Positively-Split-Ewald (PSE) method \cite{fioreRapidSamplingStochastic2017, perezpelaezComplexFluidsGpu2022, fishLibMobilityPythonLibrary2026} for a fast computation of its action $\bm{\mathcal{M}} \cdot(\bullet)$ on a given vector. Once $\bm{U}$ is computed, the positions of the markers are updated by a temporal integration of Eq.\ \eqref{eq: mob_problem_settlin_in_an_array}; we use an implicit scheme based on a backward differentiation formula (BDF) with adaptive time stepping \cite{brownVODEVariableCoefficientODE1989} -- to alleviate the numerical stiffness arising from the elastic forces. 

In our simulations, the fiber is initially straight and oriented perpendicularly to the direction of gravity. It is positioned halfway the depth of the computational domain, at $D_O/2$, with its midpoint at a given horizontal and vertical distance, $D_x$ and $D_z$, from the center (see Fig.\ \ref{fig: problemDescriptionSettlingInAnArrayOfPillars}$(b)$). 

\begin{figure}[h]
   \centering\includegraphics[width=0.5\textwidth]{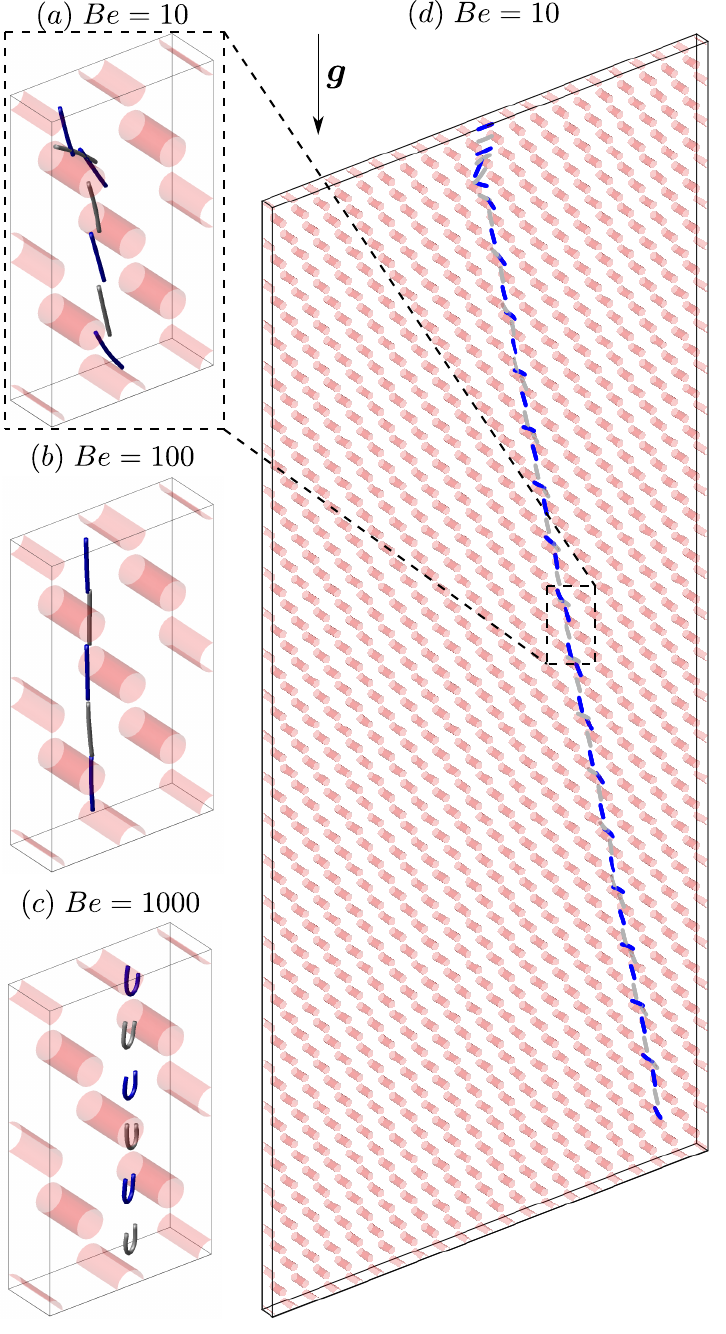}
    \caption{Chronophotographs of a flexible fiber settling through an array of pillars. The time step between two consecutive frames (shown with alternative colors: blue and gray) is $\Delta t = 3T$, where $T=L\eta/W$ is the characteristic settling time. The snapshots are taken at long times for a fixed value of the relative length, $\xi=2.36$, and various values of the elastogravitational number: $(a)\;Be=10$, $(b)\;Be=100$, and $(c)\;Be=1000$ (see also Movies 1, 3, and 4 in the Supplemental Material \cite{SSM}). The panel $(d)$ shows the unwrapped trajectory of the fiber for $Be=10$ (see also Movie 2 in the Supplemental Material \cite{SSM}). For the latter, the considered time interval is $0\leq t/T \leq 300$.
    }
    \label{fig: fiberTrajectorySettlingInAnArray}
\end{figure}

\section{Results and discussion}
\label{sec: results_and_discussion}
 The dynamics of a single flexible fiber settling under gravity in a quiescent viscous fluid embedded with a rigid obstacle have been described and analyzed in detail in our previous study \cite{makangaObstacleinducedLateralDispersion2023}. The resulting findings provide a basis to investigate the dynamics in the present work. While settling against an obstacle of arbitrary shape, a flexible fiber can either glide or remain trapped around it. These two outcomes result from the complex interplay between internal elastic stresses, contact forces, and hydrodynamic interactions. The trapping states lead to nontrivial conformations of the fiber that result from the balance between its elastic response, gravity, and friction on the surface of the obstacle. While in the gliding case, the fiber experiences a drift motion that leads to a lateral displacement. The magnitude of the latter depends on the mechanical and geometric properties of the fiber, as well as the obstacle-induced orientation.
 
 In an array of rigid pillars, the lattice arrangement and spacing may affect the lateral displacement of the fiber and, consequently, its migration. In the following, we explore the two-dimensional parameter space ($Be,\; \xi$), $10\leq Be \leq 1000$ and $2.36 \leq \xi = L/\omega \leq 6.28$, for a given lattice spacing $d/\omega \approx 3.53$ and depth $D_O/\omega \approx 2.36$. For each simulation, the fiber is released at $D_x/\omega \approx 0.12$ and $D_z/\omega\approx 5.22$, from the center of the computational domain.

\begin{figure}[h]
   \centering\includegraphics[width=0.7\textwidth]{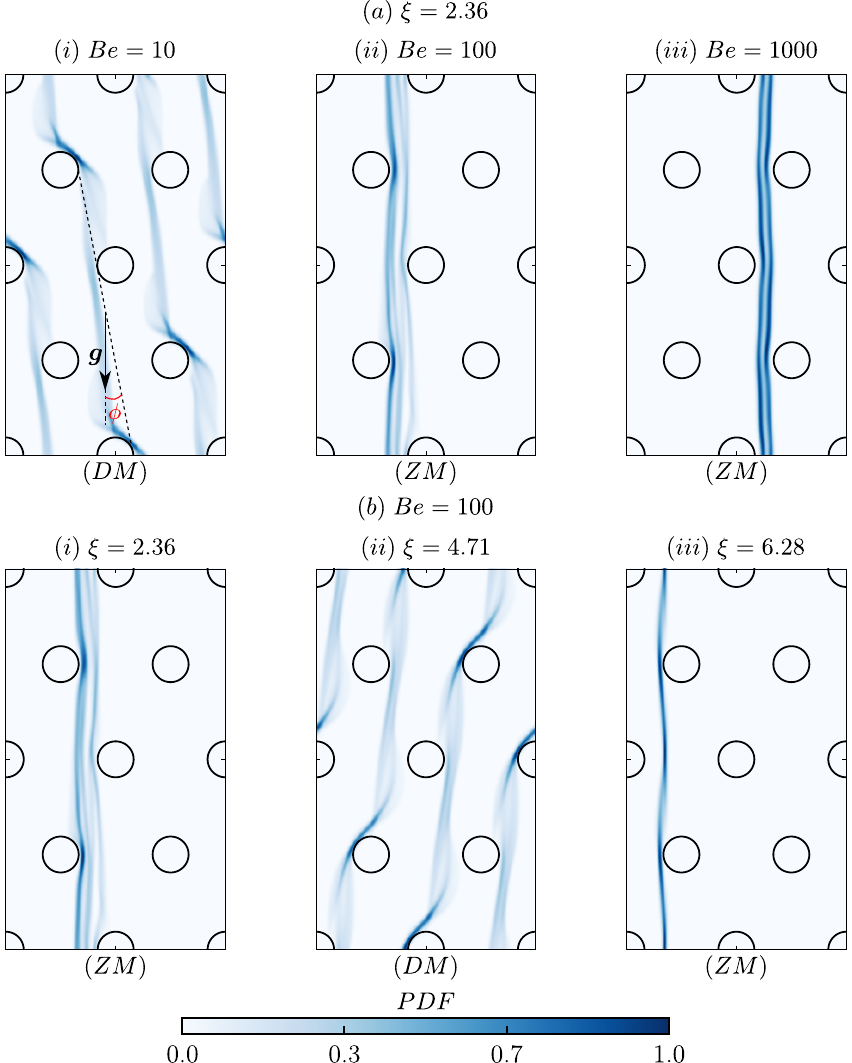}
    \caption{Probability distribution of the fiber centerline in the array of pillars at long times $(100 \leq t/T \leq 300)$. These results are obtained from the explored two-dimensional parameter space, $10\leq Be \leq 1000$ and $2.36 \leq \xi \leq 6.28$. ZM and DM stand for “Zigzag Mode" and “Displacement Mode", respectively. $\phi$ is the long-time migration angle of the fiber through the array.}
    \label{fig: MigrationModesSettlingInAnArrayOfPillars}
\end{figure}

\subsection{Long-time dynamics: ``zigzag" and ``displacement" modes}

 We start by characterizing the long-time dynamics of the fiber through the array. The considered time interval is $100\leq t/T \leq 300$, where $T=L\eta/W$ is the characteristic settling time. Figure \ref{fig: fiberTrajectorySettlingInAnArray} shows the chronophotographs, i.e., time sequences, of the fiber dynamics for a fixed relative length, $\xi=2.36$, and various values of the elastogravitational number, $10\leq Be \leq 1000$. In the stiff limit ($Be=10$), the fiber experiences a net displacement relative to the direction of gravity, i.e., lateral displacement (see Fig.\ \ref{fig: fiberTrajectorySettlingInAnArray}$(a)$, see also Movie 1 in the Supplemental Material \cite{SSM}). The corresponding unwrapped trajectory is shown in Fig.\ \ref{fig: fiberTrajectorySettlingInAnArray}$(d)$, see also Movie 2 in the Supplemental Material \cite{SSM}. In the semi-flexible ($Be=100$) and flexible ($Be=1000$) regimes, there is no average displacement relative to the direction of gravity, the fiber exhibits a slalom motion through the array (see Figs.\ \ref{fig: fiberTrajectorySettlingInAnArray}$(b)$ and \ref{fig: fiberTrajectorySettlingInAnArray}$(c)$, see also Movies 3 and 4 in the Supplemental Material \cite{SSM}). We refer to the two observed migration modes -- whose underlying mechanisms will be detailed in Sec. \ref{sec:scattering_induced_by_fiber_pillar_interactions} -- respectively as the “displacement" and “zigzag" modes.

\begin{figure}[h]
   \centering
    \includegraphics[width=0.8\textwidth]{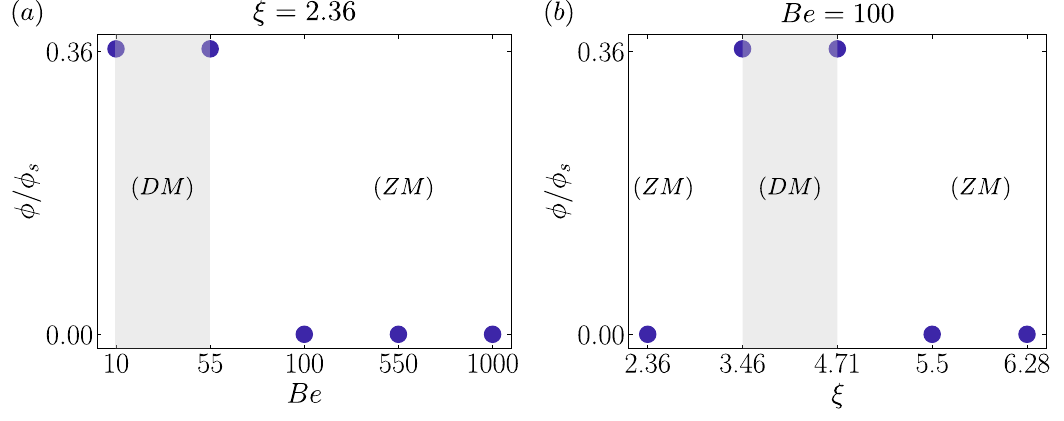}
    \caption{Migration angle relative to the structural angle of the array, $\phi/\phi_s$. $(a)$ As function of the elastogravitational number $Be$ $(10\leq Be \leq 100)$ for a fixed value of the relative fiber length $\xi=2.36$. $(b)$  As function of the relative fiber  length $(2.36\leq \xi \leq 6.28)$ for a fixed value of the elastogravitational number $Be=100$. ZM and DM stand for “Zigzag Mode" and “Displacement Mode", respectively.} 
    \label{fig: normalizedMigrationAngleVsBeAndXi}
\end{figure}

In order to characterize the trajectory of the fiber in these migration modes, we compute the probability density function (PDF) of finding the fiber centerline in the array of pillars at long times, see Fig.\ \ref{fig: MigrationModesSettlingInAnArrayOfPillars}. Note that, for the sake of clarity, only representative cases from the explored parameter space are illustrated. The two migration modes, “zigzag" and “displacement", are respectively denoted as ZM and DM in Fig.\ \ref{fig: MigrationModesSettlingInAnArrayOfPillars}. We denote by $\phi$ the migration angle of the fiber through the array (see Fig.\ \ref{fig: MigrationModesSettlingInAnArrayOfPillars}$(a)$, panel $(i)$).  In the zigzag mode (see Fig.\ \ref{fig: MigrationModesSettlingInAnArrayOfPillars}$(a)$, panels $(ii)\;\text{and}\;(iii)$; Fig.\ \ref{fig: MigrationModesSettlingInAnArrayOfPillars}$(b)$, panels $(i)\;\text{and}\;(iii)$), the trajectory of the fiber falls in a quasi-straight line with a migration angle close to zero, $\phi \approx 0$. In contrast, in the displacement mode (see Fig.\ \ref{fig: MigrationModesSettlingInAnArrayOfPillars}$(a)$, panel $(i)$; Fig.\ \ref{fig: MigrationModesSettlingInAnArrayOfPillars}$(b)$, panel $(ii)$), the trajectory of the fiber falls in a cyclical skew-bumping path with a cyclic period $n'=3$, i.e., the number of rows of pillars after which the bumping pattern repeats itself. The resulting nonzero migration angle is thus given by $\tan{(\phi)}\approx 1/(n'n \alpha_z)$, and can be related to the structural angle of the array, $\tan{(\phi)}\approx \tan{(\phi_s)}/n'$. The latter relation implies that $\phi < \phi_s$, thus the fiber is always traveling with an angle that is less than the structural angle of the array. This finding is illustrated in Fig.\ \ref{fig: normalizedMigrationAngleVsBeAndXi}, which shows the migration angle relative to the structural angle, $\phi/\phi_s$, as function of the elastogravitational number $Be$ $(10 \leq Be \leq 100)$ for a fixed value of the relative length $\xi=2.36$ (Fig.\ \ref{fig: normalizedMigrationAngleVsBeAndXi}$(a)$), and as function of the relative length $\xi$ $(2.36 \leq \xi \leq 6.28)$ for a fixed value of the elastogravitational number $Be=100$ (Fig.\ \ref{fig: normalizedMigrationAngleVsBeAndXi}$(b)$). Furthermore, the migration angle is found to be the same, $\phi \approx 0.36\phi_s$, for all displacement modes, see the gray-shaded regions in Fig.\ \ref{fig: normalizedMigrationAngleVsBeAndXi}. Therefore, for a given lattice arrangement, when its trajectory does not fall in a quasi-straight line at long times, the fiber will travel laterally with a constant angle regardless of its mechanical and geometric properties.

The above results stem from a fiber that is initially straight and oriented perpendicularly to the direction of gravity. In Appendix \ref{app: effects_initial_orientation_migration}, we report results for various initial orientations. We find that the migration modes are not sensitive to the latter, provided the fiber is not initially perfectly aligned with gravity. 

Now that we have characterized the long-time trajectory of the fiber through the array, we discuss, in the next section, the mechanisms at play at short times $(0 \leq t/T < 100)$ that lead to the zigzag and displacement modes.

\subsection{Short-time dynamics: scattering induced by fiber-pillar interactions}
\label{sec:scattering_induced_by_fiber_pillar_interactions}

The short-time dynamics of the fiber is governed by repetitive steric collisions and hydrodynamic interactions with nearby pillars. As a result, the fiber experiences a scattering process, which determines its migration mode at long times. This scattering process can be decomposed into three steps: (i) migration toward a given pillar of the array, (ii) collision followed by gliding along the surface of the pillar, and (iii) escape.

\subsubsection{Analysis of the scattering events}

\begin{figure}[h]
   \centering
    \includegraphics[width=0.6\textwidth]{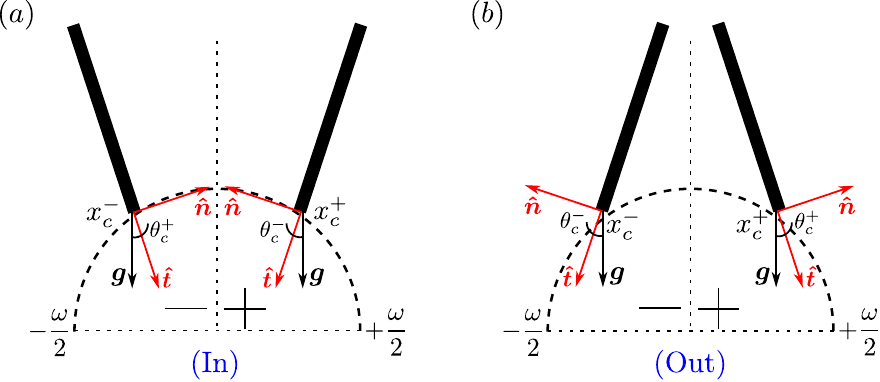}
    \caption{Schematic of the two scattering modes induced by fiber-pillar interactions, for which we define qualitative features by analogy with electrostatic interactions. $(a)$ In-scattering mode: the contact angle, $\theta_c$, and the $x-$component of the contact point, $x_c$, are unlike-charged, $(\theta_c^+, x_c^-)$ or $(\theta_c^-, x_c^+)$. $(b)$ Out-scattering mode: the contact angle, $\theta_c$, and the $x-$component of the contact point, $x_c$, are like-charged, $(\theta_c^+, x_c^+)$ or $(\theta_c^-, x_c^-)$. $x_c$ is defined with respect to the center of the pillar, and $\theta_c$ measures the orientation of the unit tangent vector at the fiber tip, $\bm{\hat{t}}$, with respect to gravity, $\bm{g}$.} 
    \label{fig: positiveNegativeScatteringSettlingInAnArray}
\end{figure}

We denote by $x_c$ the $x-$component of the contact point, with respect to the center of the pillar, between the fiber tip and the surface of the pillar; and by $\theta_c$, the oriented contact angle defined between the direction of gravity and the tangent vector $\bm{\hat{t}}$ at the fiber tip. Figure\ \ref{fig: positiveNegativeScatteringSettlingInAnArray} shows the two scattering modes induced by fiber-pillar interactions, for which we define qualitative features by analogy with electrostatic interactions:
\begin{itemize}
\item \textit{In-scattering}: $\theta_c$ and $x_c$ are unlike-charged, i.e., $(\theta_c^{+}, x_c^{-})$ or $(\theta_c^{-}, x_c^{+})$, referred to as “In" events; see Fig.\ \ref{fig: positiveNegativeScatteringSettlingInAnArray}$(a)$. 
\item \textit{Out-scattering}: $\theta_c$ and $x_c$ are like-charged, i.e., $(\theta_c^{+}, x_c^{+})$ or $(\theta_c^{-}, x_c^{-})$, referred to as “Out" events; see Fig.\ \ref{fig: positiveNegativeScatteringSettlingInAnArray}$(b)$. 
\end{itemize}

\begin{figure}[htb!]
   \centering
    \includegraphics[width=\textwidth]{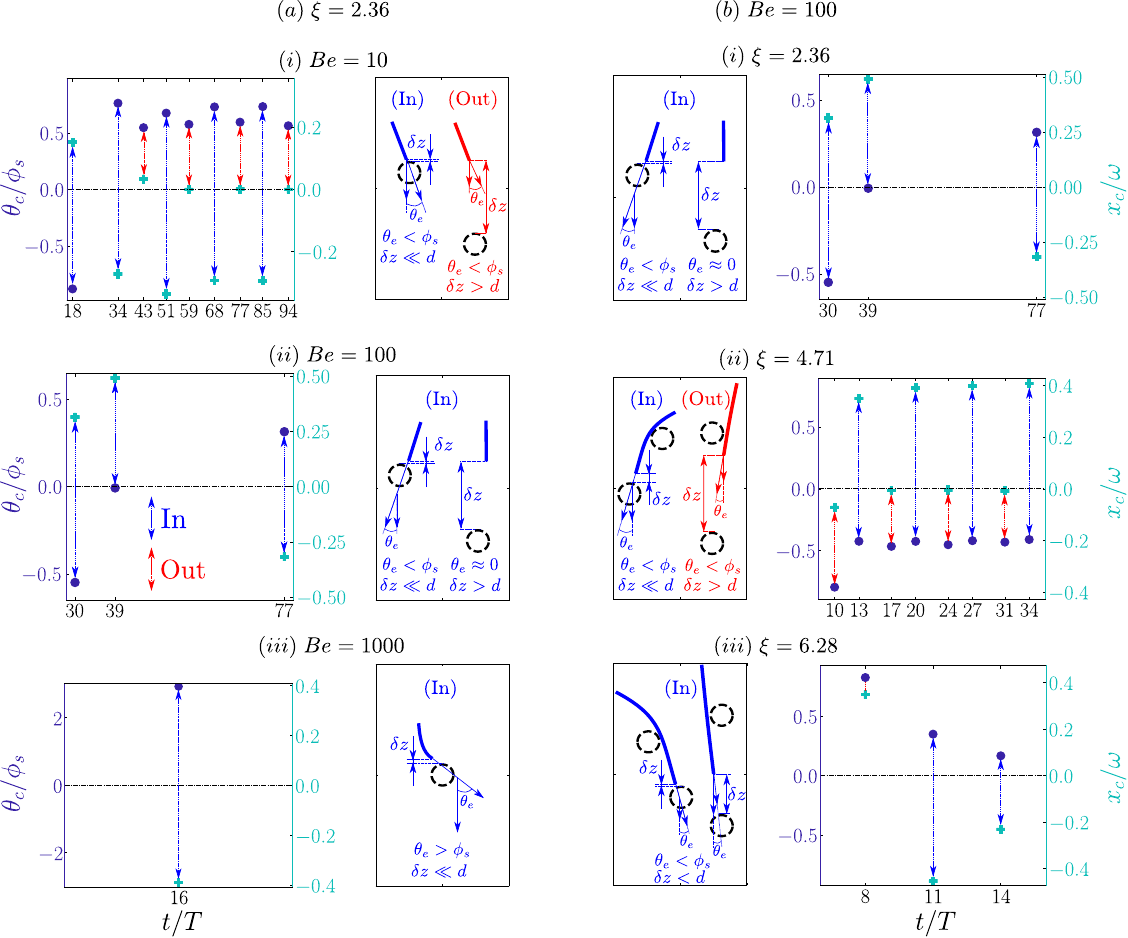}
    \caption{Results from numerical simulations showing the scattering process at short times (left and right panels in $(a)$ and $(b)$, respectively) and illustration of typical scattering modes (right and left panels in $(a)$ and $(b)$, respectively). $(a)$ As function of the elastogravitational number $Be$ $(10 \leq Be \leq 1000)$, for a fixed value of the relative fiber length $\xi=2.36$. $(b)$ As function of the relative fiber length $\xi$ $(2.36 \leq \xi \leq 6.28)$, for a fixed value of the elastogravitational number $Be=100$. In the scattering process, $\theta_c/\phi_s$ and $x_c/\omega$ are the contact angle (scaled by the structural angle of the array, $\phi_s$) and the contact point (scaled by the width of the pillars, $\omega$) between the fiber tip and the surface of the pillar, respectively. In the illustration of the scattering modes, $\theta_e$ and $d$ are the escape angle, i.e., the angle at which the fiber escapes from the previous pillar before collision, and the center-to-center distance between two adjacent pillars, respectively.} 
    \label{fig: illustrationOfInAndOutEvents}
\end{figure}

With these pictures in mind, we now proceed to discuss the features of the scattering process. To do so, we track the fiber tip that collides with a given pillar in the array at short times $(0 \leq t/T < 100)$, and we describe the different scattering modes that occur at each collision event. These results are shown in Fig.\ \ref{fig: illustrationOfInAndOutEvents} for the explored parameter space.\\

We first investigate the effects of the elastogravitational number $(10\leq Be \leq 1000)$ for a fixed value of the relative length ($\xi=2.36$), see Fig.\ \ref{fig: illustrationOfInAndOutEvents}$(a)$. In the stiff limit ($Be=10$), after its first collision, which results from its initial configuration, the fiber follows a scattering process, see left panel $(i)$ in  Fig.\ \ref{fig: illustrationOfInAndOutEvents}$(a)$ (see also Movie 1 in the Supplemental Material \cite{SSM}). The fiber first falls two times consecutively in the In-scattering mode, then alternates in a cyclic manner between Out- and In-scattering modes. This cyclic scattering pattern is a signature of a migration in a displacement mode at long times (see Fig.\ \ref{fig: MigrationModesSettlingInAnArrayOfPillars}$(a)$, panel $(i)$). Owing to its stiffness, the fiber drifts sideways in a skew configuration after escaping from the surface of the pillar with which it collided. The subsequent lateral displacement increases with the settling distance \cite{makangaObstacleinducedLateralDispersion2023}.
When the vertical distance that the fiber has to settle is less than the center-to-center distance between two adjacent pillars, $\delta z \ll d$, and its escape angle, i.e., the angle at which the fiber escapes from the previous pillar
before collision, is less than the structural angle of the array, $\theta_e < \phi_s$ (see Fig.\ \ref{fig: illustrationOfInAndOutEvents}$(a)$, right panel $(i)$); the fiber undergoes a weak lateral displacement compared to its length, $\delta x / L \ll 1$. As a result, its contact angle is close to its escape angle, $\theta_c \sim \theta_e$. Therefore, the resulting collision events occur with positive contact angles on the left side of the collided pillars $(\theta_c^+, x_c^-)$, thereby leading to In-scattering events. Conversely, when $\delta z > d$ (see Fig.\ \ref{fig: illustrationOfInAndOutEvents}$(a)$, right panel $(i)$), the fiber undergoes a larger lateral displacement and has more time to reorient than in In-scattering events. As a result, its contact angle is larger than its escape angle, $\theta_c > \theta_e$. Therefore, the resulting collision events occur with positive contact angles on the right side of the collided pillars $(\theta_c^+, x_c^+)$, thereby leading to Out-scattering events. In the flexible regime ($Be=1000$), due to its short elastic relaxation timescale, $\mathcal{O}(Be^{-1}) \ll 1$, the fiber escapes from its first collision with a hook-like shape and a positive escape angle (see also Movie 4 in the Supplemental Material \cite{SSM}). Since the allowed vertical distance $\delta z$ for the fiber to settle until collision is smaller than $d$ (see Fig.\ \ref{fig: illustrationOfInAndOutEvents}$(a)$, right panel $(iii)$), it undergoes a weak lateral displacement with respect to its length, $\delta x/L \ll 1$. As a result, its contact angle is close to its escape angle, $\theta_c \sim \theta_e$. Therefore, the resulting collision event occurs with a positive contact angle on the left side of the collided pillar, $(\theta_c^+, x_c^-)$. This leads to the In-scattering event shown in the scattering process (see Fig.\ \ref{fig: illustrationOfInAndOutEvents}$(a)$, left panel $(iii)$). The fiber then adopts a quasi-symmetric U-like shape, with its end-to-end width smaller than the center-to-center distance between two adjacent pillars. The trajectory of the fiber therefore falls in a quasi-straight line with no average displacement relative to the direction of gravity (see Fig.\ \ref{fig: MigrationModesSettlingInAnArrayOfPillars}$(a)$, panel $(iii)$). Finally, in the semi-flexible regime ($Be=100$), the fiber tip tends to align with gravity as it escapes from the pillars (see Fig.\ \ref{fig: illustrationOfInAndOutEvents}$(a)$, right panel $(ii)$); consequently, the fiber exhibits weak lateral displacements, $\delta x /L \ll 1$. As a result, all the scattering modes fall in “In" events, see left panel $(ii)$ in  Fig.\ \ref{fig: illustrationOfInAndOutEvents}$(a)$ (see also Movie 3 in the Supplemental Material \cite{SSM}).
Thus, the trajectory of the fiber, as in the flexible regime, falls at long times in a quasi-straight line  (see Fig.\ \ref{fig: MigrationModesSettlingInAnArrayOfPillars}$(a)$, panel $(ii)$).\\

On the other hand, the relative length  $(2.36 \leq \xi \leq 6.28)$  affects -- in the semi-flexible regime $(Be=100)$ -- the short-time dynamics as follows (see Fig.\ \ref{fig: illustrationOfInAndOutEvents}$(b)$): for $\xi=2.36$, the fiber length is smaller than the lattice spacing, $L/d \approx 0.67$. All the scattering modes fall in “In" events (Fig.\ \ref{fig: illustrationOfInAndOutEvents}$(b)$, panel $(i)$), the surrounding pillars prevent the fiber from undergoing lateral displacements (see also Movie 3 in the Supplemental Material \cite{SSM}). As a result, the fiber migrates through the array in a quasi-straight line (Fig.\ \ref{fig: MigrationModesSettlingInAnArrayOfPillars}$(b)$, panel $(i)$). When the fiber length is larger than the lattice spacing, $L/d > 1$, after its first collision, the fiber cannot align with the direction of gravity while escaping from the pillar (see also Movies 5 and 6 in the Supplemental Material \cite{SSM}). As a result, the fiber settles a vertical distance less than $d$, and its tip collides with the subsequent pillar with a contact angle close to $\phi_s$. This leads to the two Out-scattering events shown at $t/T=10$ and $t/T=8$ in Fig.\ \ref{fig: illustrationOfInAndOutEvents}$(b)$ (right panels $(ii)$ and $(iii)$), respectively for $\xi=4.71$ and $\xi=6.28$. For each of these Out-scattering events, the fiber bends significantly while escaping from the pillar, such that its lower tip tends to align with the direction of gravity. Therefore, its escape angle becomes smaller than the structural angle of the array, $\theta_e < \phi_s$ (see Fig.\ \ref{fig: illustrationOfInAndOutEvents}$(b)$, left panels $(ii)$ and $(iii)$). For $\xi=4.71$, the fiber alternates in a cyclic manner between In- and Out-scattering events respectively when $\delta z \ll d$ and  $\delta z > d$ (see Fig.\ \ref{fig: illustrationOfInAndOutEvents}$(b)$, right panel $(ii)$). Thus, the trajectory of the fiber falls at long times in a cyclical skew-bumping path (see Fig.\ \ref{fig: MigrationModesSettlingInAnArrayOfPillars}$(b)$, panel $(ii)$). Conversely, for $\xi=6.28$, owing to its significant length, $L/d \approx 2$, the vertical distance that the fiber has to settle is always smaller than the center-to-center distance between two adjacent pillars, $\delta z < d$. As a result, all subsequent scattering modes fall in “In" events (see Fig.\ \ref{fig: illustrationOfInAndOutEvents}$(b)$, right panel $(iii)$). Thus, the trajectory of the fiber falls at long times in a quasi-straight line (see Fig.\ \ref{fig: MigrationModesSettlingInAnArrayOfPillars}$(b)$, panel $(iii)$).\\

\subsubsection{Theoretical prediction of the scattering events}

Now that we have a clear understanding of how the scattering process leads to the two migration modes, we use, in the following, a theoretical model to predict in which migration mode the trajectory of the fiber will fall at long times, owing to the geometry of the array, i.e., lattice arrangement and spacing.

\begin{figure}[h]
   \centering   
    \includegraphics[width=0.81\textwidth]{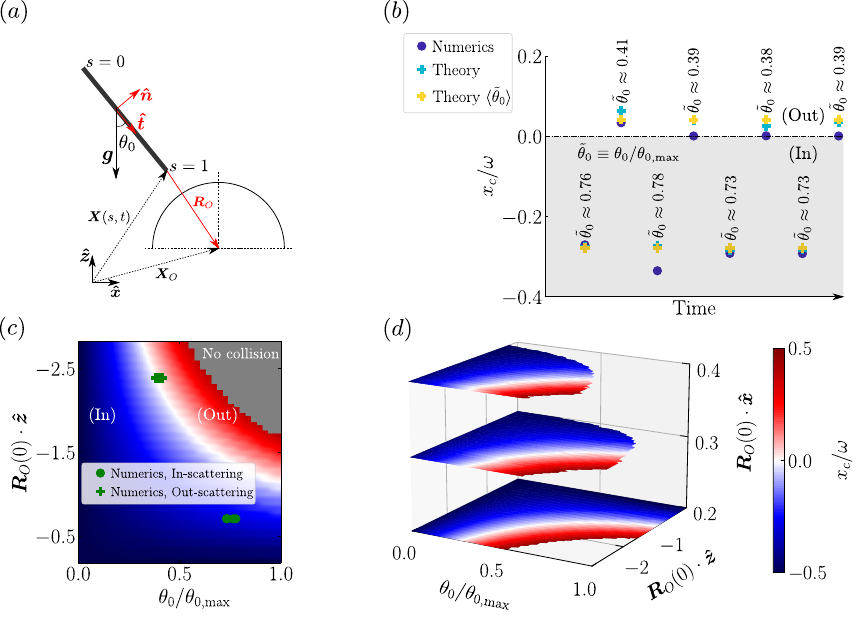}
    \caption{Theoretical investigation for $Be=10$ and $\xi=2.36$. $(a)$ Illustration of the migration of the fiber toward the pillar. The position of the fiber centerline, $\bm{X}(s,t)$, is parameterized by the arc length coordinate $s \in [0, 1]$, and the material frame $(\bm{\hat{t}}, \bm{\hat{n}})$ is attached to the former. The orientation angle $\theta_0$, at the midpoint, is defined between the unit tangent $\bm{\hat{t}}$ and the direction of gravity -$\bm{\hat{z}}$. The center of the pillar $\bm{X}_O$ is initially positioned at a given distance from the lower end of the fiber, $\bm{R}_O(0)=\bm{X}_O - \bm{X}(1,0)$. $(b)$ Time evolution of the scaled contact points, $x_c/\omega$. Comparison between the results from numerical simulations (circle markers) and theoretical predictions (cross markers). $(c)$ Phase diagram showing the scaled contact points, $x_c/\omega$, for $\bm{R}_O(0)\cdot \bm{\hat{x}}=0.2$,  $0.2 \leq |\bm{R}_O(0)\cdot \bm{\hat{z}}|\leq 2.8$, and $0\leq \theta/\theta_{0,\text{max}} \leq 1$, where $\theta_{0,\text{max}}=\pi/6$. The gray-shaded region and markers correspond to non-collision events and numerical simulation results, respectively. $(d)$ Effects of $\bm{R}_O(0)\cdot \bm{\hat{x}}$, $0.2 \leq \bm{R}_O(0)\cdot \bm{\hat{x}} \leq 0.4$, on the phase diagram shown in panel $(c)$.}
    \label{fig: theoreticalInvestigationSettlingInAnArray}
\end{figure}

We first derive the equations governing the trajectory of a settling fiber in an obstacle-free domain. We denote by $\bm{X}(s,t)$ the space curve, which represents the position of the centerline of the fiber at a given time $t$, parameterized by the arc length coordinate $s \in [0, L]$, see Fig.\ \ref{fig: theoreticalInvestigationSettlingInAnArray}$(a)$. In this section, unless otherwise stated, we scale lengths upon $L$ and time upon $T$: $\bm{X} = L\widetilde{\bm{X}}$, $s = L\tilde{s}$, and $t=T\tilde{t}$; and we drop the tilde for notation convenience.

In the limit of small deformation, the position of the lower end of the fiber, $\bm{X}(1,t)$, can be written as 

\begin{equation}
\bm{X}(1,t) = \bm{X}(1/2,t) + \frac{1}{2}\bm{\hat{t}}(\theta(t)),
\label{eq: position_end_point}
\end{equation}
where $\bm{X}(1/2,t)$ and $\bm{\hat{t}}=\partial_s \bm{X}(s,t)|_{s=1/2}$ are the position and the unit tangent at the midpoint of the fiber, respectively; and $\theta$ measures the orientation of the fiber with respect to the direction of gravity, $-\bm{\hat{z}}$. The velocity of the lower end of the fiber is thus obtained by differentiating Eq.\ \eqref{eq: position_end_point} with respect to time,

\begin{equation}
\frac{\partial \bm{X}(1,t) }{\partial t } \equiv \bm{U} =  \overline{\bm{U}} + \frac{1}{2}\frac{\partial \theta (t)}{\partial t}\frac{\partial \bm{\hat{t}}(\theta)}{\partial \theta},
\label{eq: velocity_end_point}
\end{equation}
where $\overline{\bm{U}} \equiv \partial_t \bm{X}(1/2,t)$ is the fiber midpoint velocity. The latter is given by Li \emph{et al.} \cite{liSedimentationFlexibleFilaments2013a}, $\overline{\bm{U}} = 2c_0 \cos{(\theta)}\bm{\hat{t}}(\theta) - c_0 \sin{(\theta)}\bm{\hat{n}}(\theta)$, where $\theta$ satisfies the following differential equation:

\begin{equation}
    \frac{\partial \theta}{\partial t} = CBe (\sin{(\theta)}),
    \label{eq: orientation_angle}
\end{equation}
with 

\begin{equation}
    C = \frac{7}{400} + c_0^{-1}\left(\frac{1813 - 300\pi^2 + 630\ln{(2)}}{18000}\right),
    \label{eq: constant_big_c}
\end{equation}
where $c_0 = \ln{(1/\varepsilon^2 e)}$ with $\varepsilon=2a/L$ being the aspect ratio of the fiber. Thus, the components of $\bm{U}$ in the laboratory frame, $U_x$ and $U_z$, are obtained by dotting Eq.\ \eqref{eq: velocity_end_point} respectively with the unit vectors $\bm{\hat{x}}$ and $\bm{\hat{z}}$,

\begin{equation}
\left\{
\begin{aligned}
U_x &= \bm{U}\cdot \bm{\hat{x}} =  c_0\left(\frac{ h}{1 + h^2}\right) +  CBe\left(\frac{h}{(1 + h^2)^{3/2}}\right)\\\\
U_z &= \bm{U}\cdot \bm{\hat{z}} = -c_0\left(1 + \frac{1}{1 + h^2}\right) + CBe\left(\frac{h^2}{(1 + h^2)^{3/2}}\right)
\end{aligned}
\right.,
\label{eq: velocity_end_point_reference_frame}
\end{equation} 
where

\begin{equation}
    h(t) = \tan{(\theta_0)}\exp{[(2CBe)t]},
    \label{eq: expression_of_h}
\end{equation}
with $\theta_0 = \theta(0)$. Then, writing $\bm{X}(1,t)=X(t)\bm{\hat{x}} + Z(t)\bm{\hat{z}}$ and assuming that the fiber is initially centered at the origin, i.e., $X(0)=\sin{(\theta_0)}/2$ and $Z(0)=-\cos{(\theta_0)}/2$, the time-dependent parametric equation of the path $\mathcal{\bm{L}}$ -- described by the lower end of the fiber -- is obtained by integrating Eq.\ \eqref{eq: velocity_end_point_reference_frame} in time,

\begin{equation}
\mathcal{\bm{L}}:
\left\{
\begin{aligned}
 & \tan{\left[\frac{CBe}{c_0}\left(2X - \frac{\tan{(\theta_0)}\exp{[(2CBe)t]}}{(1 + \tan^2{(\theta_0)}\exp{[(4CBe)t]})^{1/2}}\right) + \theta_0\right]} = \tan{(\theta_0)}\exp{[(2CBe)t]}\\\\
& Z = -c_0t - \frac{c_0}{4CBe}\ln{\left\{\frac{[1 + \tan^2{(\theta_0)}]\exp{[(4CBe)t]}}{1 + \tan^2{(\theta_0)}\exp{[(4CBe)t]}}  \right\}} - \frac{1}{2(1 + \tan^2{(\theta_0)}\exp{[(4CBe)t]})^{1/2}}
\end{aligned} \quad t \in [0, +\infty).
\right.
\label{eq: paramteric_equations_trajectory_midpoint}
\end{equation} 

Then, to account for the effects of the obstacles on $\mathcal{L}$, we consider a dilute array of pillars, i.e., the velocity disturbance induced by a given pillar in the vicinity of another is small in magnitude. Furthermore, we assume that the lattice spacing is much larger than the fiber length $(d/L\gg 1)$, i.e., the fiber cannot be in contact with more than one pillar during a collision event. To characterize a collision event, we define the initial orientation of the fiber as its escape angle, $\theta_0 \equiv \theta_e$, and we denote by $\bm{R}_O(t) = \bm{X}_O - \bm{X}(1,t)$, the relative position of the center of a given pillar of the array from the lower end of the fiber, see Fig.\ \ref{fig: theoreticalInvestigationSettlingInAnArray}$(a)$. Thus, for a given $\theta_0$, the contact point $(x_c, z_c)$, between the lower end of the fiber and the surface of a given pillar, is any point $(X, Z)$ of $\mathcal{\bm{L}}$ that lies on the latter. In other words, given the initial conditions, $\theta_0$ and $\bm{R}_O(0)$, the  contact point  is obtained by solving the following constrained problem numerically:

\begin{equation}
    \left\{
    \begin{aligned}
        \text{find} \quad &(X, Z) \in \mathcal{\bm{L}}\\
        \text{such that} \quad &\lVert\bm{R}_O\rVert = \frac{\omega}{2L}
    \end{aligned}
    \right..
\end{equation}

Figure \ref{fig: theoreticalInvestigationSettlingInAnArray}$(b)$ shows the time evolution of $x_c$ for $Be=10$ and $\xi=2.36$  ($\varepsilon^{-1}=15$). The results from numerical simulations and theoretical predictions correspond to the blue circle markers and cyan cross markers, respectively. 
The initial position $\bm{R}_O(0)$ and orientation $\Tilde{\theta}_0 \equiv \theta_0/\phi_s$ are taken from numerical simulations for the two scattering modes, where  $\bm{R}_O(0)=0.2\bm{\hat{x}} - 0.7\bm{\hat{z}}$ for In-scattering events and $\bm{R}_O(0)=0.2\bm{\hat{x}} - 2.4\bm{\hat{z}}$ for Out-scattering events. Although we observe small discrepancies for some values of $x_c$, probably due to the fact that our theoretical model does not take into account hydrodynamic interactions with the pillar, we find overall a good agreement between both numerical and theoretical results. This finding confirms the adequacy of our simple theoretical model to predict the scattering process that the fiber will experience, in the limit of small deformations and large lattice spacings. Since the reported values of $\tilde{\theta}_0$ vary only slightly over time, implying that the fiber escapes from pillar surfaces at the same angle ($\tilde{\theta}_0 \approx 0.39$ for Out-scattering events and $\tilde{\theta}_0 \approx 0.75$ for In-scattering events), the time average of the escape angles, $\langle \tilde{\theta}_0 \rangle$,  provides a good estimate of the initial orientation of the fiber, see the yellow cross markers in Fig.\ \ref{fig: theoreticalInvestigationSettlingInAnArray}$(b)$.       

 Then, to predict in which migration mode the trajectory of the fiber will fall at long times for a given lattice spacing, we explore the sets of contact points obtained for various $\theta_0$, $0 \leq \theta_0/\theta_{0,\mathrm{max}} \leq 1$ with $\theta_{0,\mathrm{max}}= \pi/6$, various $\bm{R}_O(0)\cdot \bm{\hat{z}}$, $0.2 \leq |\bm{R}_O(0)\cdot\bm{\hat{z}}| \leq 2.8$, various $\bm{R}_O(0)\cdot \bm{\hat{x}}$, $0.2 \leq \bm{R}_O(0)\cdot\bm{\hat{x}} \leq 0.4$, and a fixed value of the fiber length, $\xi=2.36$. Figure \ref{fig: theoreticalInvestigationSettlingInAnArray}$(c)$ shows the predicted scattering events for $Be=10$ and $\bm{R}_O(0)\cdot \bm{\hat{x}}=0.2$, in good agreement with the results from numerical simulations. We observe that the region of In-scattering events is larger than that of Out-scattering events, implying that the former are more likely to occur than the latter. In particular, a displacement can only occur if  $\bm{R}_O(0)\cdot \bm{\hat{z}} \lesssim -0.7$ and if $\theta_0$ is sufficiently large. Furthermore, the number of collision events -- with the fiber tip -- decreases as $\bm{R}_O(0)\cdot \bm{\hat{x}}$ increases, see Fig.\ \ref{fig: theoreticalInvestigationSettlingInAnArray}$(d)$. It is worth noting that the non-collision regions -- corresponding to a passage either to the left or the right of the pillar --  also encompass collision events that do not occur between the fiber tip and the surface of the pillar. Altogether, these predictions give a map of the scattering events depending on the geometric arrangement of the lattice and on the initial fiber orientation. They provide guidelines for the design of gravity-driven fiber separation devices: the vertical separation between pillars and the escape angle of the fibers after collision, which becomes the initial angle for the next pillar, must both be large enough to enable sufficient lateral drift to lead to a displacement mode.
This implies that rigid fibers  are more prone to a lateral drift and thus, to a displacement mode, unlike highly flexible fibers, whose trajectories fall into a quasi-straight line owing to their short elastic relaxation timescale.

\begin{figure}[h]
   \centering
    \includegraphics[width=\textwidth]{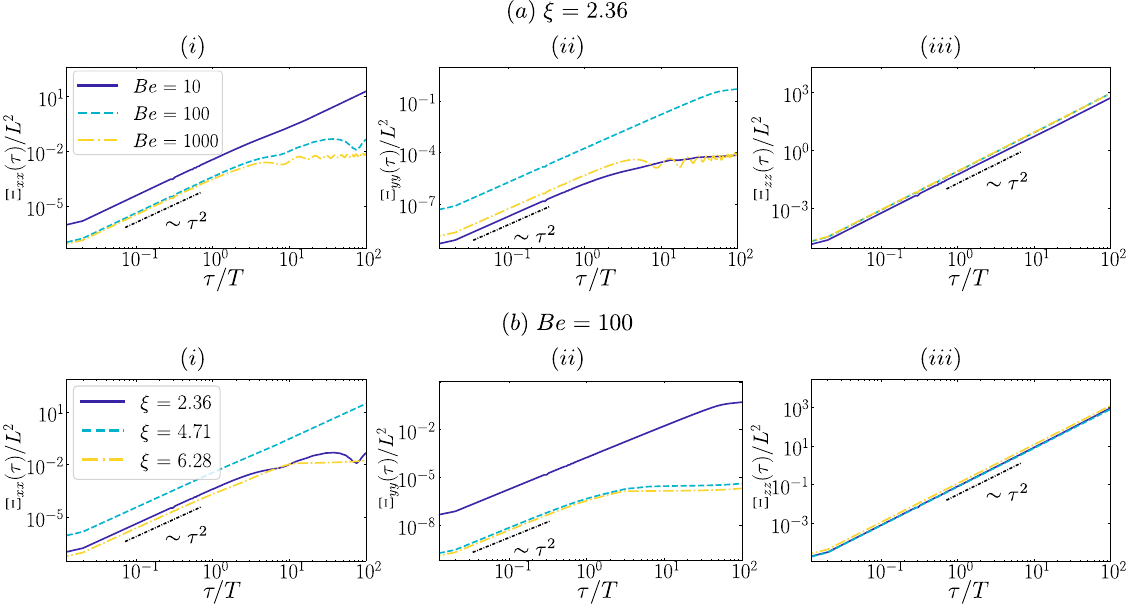}
    \caption{Results from numerical simulations showing the time evolutions of the diagonal components of the mean-squared-displacement tensor (MSD) scaled by the length of the fiber squared, $\Xi_{xx}/L^2$, $\Xi_{yy}/L^2$, and $\Xi_{zz}/L^2$. $(a)$ As function of the elastogravitational number $Be$ $(10 \leq Be \leq 1000)$, for a fixed value of the relative fiber length $\xi=2.36$. $(b)$ As function of the relative fiber length $\xi$ $(2.36 \leq \xi \leq 6.28)$, for a fixed value of the elastogravitational number $Be=100$.}
    \label{fig: meanSquareDisplacementSettlingInAnArray}
\end{figure}

\subsection{Long-time transport properties}
\label{sec:long_time_transport_properties}

We now set out to describe the long-time transport properties of the fiber. To do so, we introduce the mean- squared-displacement (MSD) tensor $\bm{\Xi} \in \mathbb{R}^{3\times 3}$,

\begin{equation}
    \bm{\Xi}(\tau) = \left < \left[\overline{\bm{r}}^{F}(t + \tau) - \overline{\bm{r}}^{F}(t) \right]\left[\overline{\bm{r}}^{F}(t + \tau) - \overline{\bm{r}}^{F}(t) \right] \right>,
    \label{eq: mean_square_displacement}
\end{equation}
where $\left < \bullet \right>$ denotes the ensemble average and $\overline{\bm{r}}^{F}$ is the center of mass of the fiber. Figure\ \ref{fig: meanSquareDisplacementSettlingInAnArray} shows the diagonal components of $\bm{\Xi}$: $\Xi_{xx}$, $\Xi_{yy}$, and $\Xi_{zz}$, for $10\leq Be \leq 1000$  and $2.36 \leq \xi \leq 6.28$.

 We start by describing the dynamics of the fiber along the vertical and transverse directions, $\Xi_{zz}$ (see panels $(iii)$ in Fig.\ \ref{fig: meanSquareDisplacementSettlingInAnArray}$(a)$ and Fig.\ \ref{fig: meanSquareDisplacementSettlingInAnArray}$(b)$) and $\Xi_{xx}$ (see panels $(i)$ in Fig.\ \ref{fig: meanSquareDisplacementSettlingInAnArray}$(a)$ and Fig.\ \ref{fig: meanSquareDisplacementSettlingInAnArray}$(b)$). $\Xi_{zz}$ is always ballistic ($\sim \tau^2$), whereas $\Xi_{xx}$ can exhibit a plateau -- a signature of a caged regime. These two regimes determine the migration mode of the fiber. The ballistic regime reflects the displacement mode ($[Be=10,\; \xi=2.36]$ and $[Be=100,\; \xi=4.71]$) through the repetitive collisions that lead to a lateral displacement of the fiber. Conversely, the caged regime reflects the zigzag mode ($[Be=100,\; \xi=2.36]$, $[Be=100,\; \xi=6.28]$, and $[Be=1000,\; \xi=2.36]$) through the formation of an imaginary tube by the surrounding pillars, which prevents the fiber from undergoing lateral displacements. 

\begin{figure}[htb!]
   \centering
    \includegraphics[width=\textwidth]{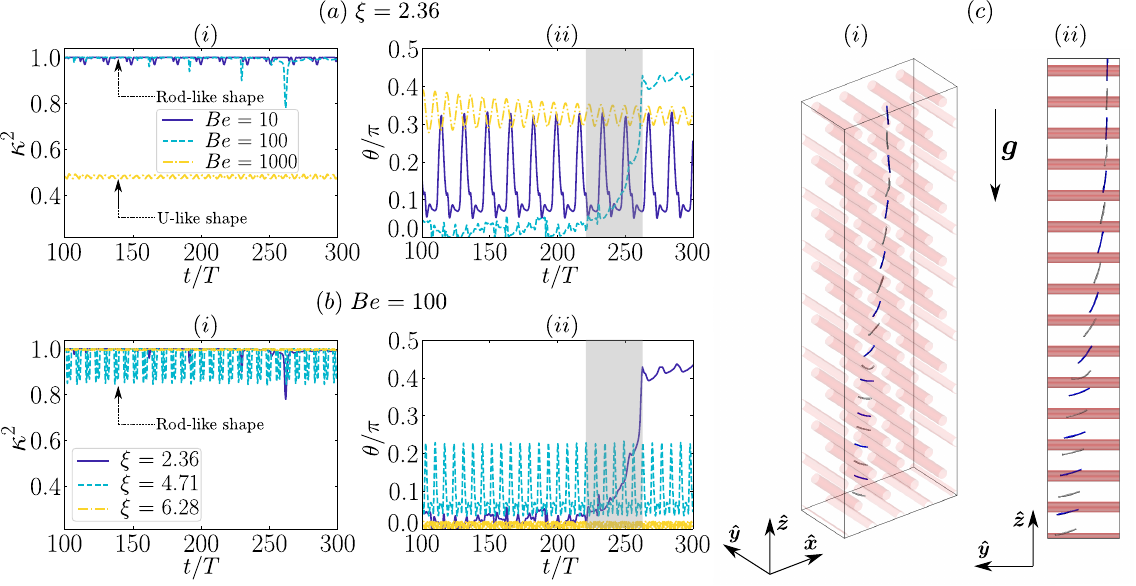}
    \caption{$(a)-(b)$ Long-time evolutions of the relative shape anisotropy $\kappa^2$ of the fiber $(i)$ and its orientation $\theta$ (measured at the fiber midpoint) with respect to the direction of gravity $(ii)$. These results have been obtained from the explored two-dimensional parameter space ($10\leq Be \leq 1000$, $2.36 \leq \xi \leq 6.28$). $(c)$ Chronophotographs showing a three-dimensional representation $(i)$ and a two-dimensional representation, in the $yz-$plane $(ii)$, of the long-time reorientation of the fiber for $Be=100$ and $\xi=2.36$, while settling through the array of pillars (see also Movie 7 in Supplemental Material \cite{SSM}). The results are shown for the time interval $215\leq t/T \leq 300$, where $T=L\eta/W$ is the characteristic settling time. The time step between two consecutive frames (shown with alternating colors: blue and gray) is $\Delta t \approx 9T$.} 
    \label{fig: orientationAndShapeVariousRigidityAndLength}
\end{figure}

 In the longitudinal direction (see panels $(ii)$ in Fig.\ \ref{fig: meanSquareDisplacementSettlingInAnArray}$(a)$ and Fig.\ \ref{fig: meanSquareDisplacementSettlingInAnArray}$(b)$), although of small magnitude compared to the vertical and transverse displacements, we observe that $\Xi_{yy}$ grows ballistically before reaching a plateau. This transition, from a ballistic to a caged regime, corresponds to a finite lateral drift of the fiber along the $y-$axis. Therefore, the motion of the fiber is not restricted to a two-dimensional plane ($xz-$plane); rather, the fiber exhibits an out-of-plane motion. To further characterize the latter, we investigate the long-time deformation and orientation of the fiber. To do so, we introduce the gyration tensor $\bm{Q} \in \mathbb{R}^{3\times3}$ \cite{blavatskaShapeAnisotropyPolymers2010, arkinGyrationTensorBased2013}, whose components are given by

\begin{equation}
    Q_{jk} = \frac{1}{N_F} \sum_{i=1}^{N_F}[r^{F}_{i, \alpha} - r^{F}_{M,\alpha}][r^{F}_{i, \beta} - r^{F}_{M,\beta}], \quad \alpha, \beta = 1, 2, 3,
    \label{eq: gyration_tensor}
\end{equation}
where $r^{F}_{i,\alpha}$ and $r^{F}_{M,\alpha}$ are respectively the components of the positions of the markers $\bm{r}^{F}_i$ and the fiber midpoint $\bm{r}^{F}_M$. Since the gyration tensor is symmetric, there exists a basis in which it can be diagonalized. Its eigenvalues, also known as \textit{principal moments}, can be combined to characterize the fiber shape. Let $(\lambda_{i})_{1 \leq i \leq 3}$ be the eigenvalues of $\bm{Q}$ and $\overline{\lambda} \equiv \mathrm{Tr}\bm{Q}/3$ their average. The relative shape anisotropy of the fiber, $\kappa^2$, is given by  

\begin{equation}
    \kappa^2 = \frac{1}{6} \sum_{i=1}^{3}\frac{(\lambda_i - \overline{\lambda})^2}{\overline{\lambda}^2} = \frac{3}{2} \frac{\mathrm{Tr}\bm{\hat{Q}}^2}{(\mathrm{Tr}\bm{Q})^2},
    \label{eq: shape_anisotropy}
\end{equation}
where $\bm{\hat{Q}} \equiv \bm{Q} - \overline{\lambda}\bm{I}$, with $\bm{I} \in \mathbb{R}^{3\times3}$ being the identity matrix. $\kappa^2$ evolves from zero to one, corresponding respectively to an isotropic and a rod-like shape. The long-time evolution of the relative shape anisotropy $\kappa^2$ and the orientation $\theta$ (measured at the fiber midpoint) are shown in Figs.\ \ref{fig: orientationAndShapeVariousRigidityAndLength}$(a)-(b)$. Both quantities are represented for various values of the elastogravitational number $Be$ $(10 \leq Be \leq 1000)$ at a fixed value of the relative length $\xi=2.36$, Fig.\ \ref{fig: orientationAndShapeVariousRigidityAndLength}$(a)$; and for various values of the relative length $\xi$ $(2.36 \leq \xi \leq 6.28)$ at a fixed value of the elastogravitational number $Be=100$, Fig.\ \ref{fig: orientationAndShapeVariousRigidityAndLength}$(b)$. In the semi-flexible regime $(10 \leq Be \leq 100)$, the fiber weakly deforms -- its shape remains quasi-straight ($\kappa^2 \approx 1$), while its orientation $\theta$ varies in the two migration modes. The time-periodic evolutions of $\theta$, with uniform amplitudes, characterize a cyclical skew-bumping trajectory, which is a signature of a migration in a displacement mode. Conversely, when $\theta \approx 0$, the fiber is quasi-parallel to the direction of gravity, thereby reflecting a migration in a zigzag mode. However for $\xi=2.36$ and $Be=100$, we observe a sudden jump in the orientation angle from $\theta \approx 0$ to $\theta \approx 0.45 \pi$ (see the gray-shaded regions in Figs.\ \ref{fig: orientationAndShapeVariousRigidityAndLength}$(a)$-$(b)$, panels $(ii)$), while the fiber is still roughly straight ($\kappa^2 \approx 1$) and executes a zigzag motion through the array. During this transition, the fiber rotates around the $x-$axis to roughly adopt its obstacle-free equilibrium shape, i.e., oriented perpendicularly to the direction of gravity \cite{marchettiDeformationFlexibleFiber2018}, in the $yz-$plane, which is the less topologically constrained direction of the array. This reorientation leads to a lateral displacement of the fiber along the $y-$axis, see Fig.\ \ref{fig: orientationAndShapeVariousRigidityAndLength}$(c)$ (see also Movie 7 in the Supplemental Material \cite{SSM}). Finally, in the highly flexible regime $(Be \geq 1000)$, the fiber adopts a roughly U-like shape ($\kappa^2 \approx 0.5$) with an average orientation angle $\theta \approx 0.35\pi$, while moving in a quasi-straight line through the array. 

\section{Conclusions}
\label{sec: conclusions}

In this paper, we have investigated the long-time behavior of a single flexible fiber settling under gravity through a periodic array of pillars in a viscous fluid. Our findings show that the long-time trajectory of the fiber falls into one of two migration modes, zigzag or displacement. In the former, there is no average displacement relative to the direction of gravity; the center of mass of the fiber follows a quasi-straight line with a migration angle close to zero. In the latter, there is a net displacement relative to the direction of gravity; the center of mass of the fiber follows a cyclical skew-bumping path.

The short-time scattering process of the fiber, which stems from repetitive collisions with the pillars, was found to govern both migration modes. In the limit of small deformations $(Be \ll 1)$ and large lattice spacings $(d/L \gg 1)$, we have proposed a theoretical model to predict the scattering process that the fiber will experience and, therefore, in which migration mode its trajectory will fall at long times. These outcomes, together with the ones obtained in our previous study \cite{makangaObstacleinducedLateralDispersion2023}, provide insights into the design of gravity-based sorting devices for the purpose of biomedical, microfluidics, and environmental applications \cite{makanga2025device}. 

Our numerical investigations have been limited to a hexagonal arrangement of the array. Since the lattice spacing affects the long-time dynamics of the fiber, it is clear that the lattice arrangement will have similar effects on its lateral displacement. For instance, to avoid the out-of-plane motions of the fiber, see Sec. \ref{sec:long_time_transport_properties}, one can use a grid-like arrangement to constrain the fiber motion to a two-dimensional plane. In practical situations such as wastewater treatment processes, suspensions of elongated particles settle against screens, i.e., arrays of long metal bars. The fibers will eventually clump together to form clusters \cite{metzgerExperimentalInvestigationInstability2007, saintillanGrowthConcentrationFluctuations2006a, manikantanEffectFlexibilityGrowth2016} which will interact hydrodynamically and collide with their surroundings. From this complex coupling, may emerge unexpected collective dynamics whose understanding will inspire future works on the transport of fiber suspensions in complex environments.

\begin{acknowledgments}
B.D.\ and U.M.\ acknowledge support from the French National Research Agency (ANR), under awards ANR- 20-CE30-0006 and ANR-25-CE30-4710. B.D.\ also thanks the NVIDIA Academic Partnership program for providing GPU hardware for performing some of the simulations reported here.
\end{acknowledgments}

\section*{Data availability}
All the data that support the findings of this study are available from the corresponding author upon reasonable request.

\section*{Code availability}
All the codes that support the findings of this study are available from the corresponding author upon reasonable request.  

\section*{Declaration of interests}
The authors report no conflict of interest.

\appendix

\section{Details on the numerical methods}
\label{app: details_on_the_numerical_method}

\subsection{Fiber model}
\begin{figure}[h]
   \centering
    \includegraphics[width=0.4\textwidth]{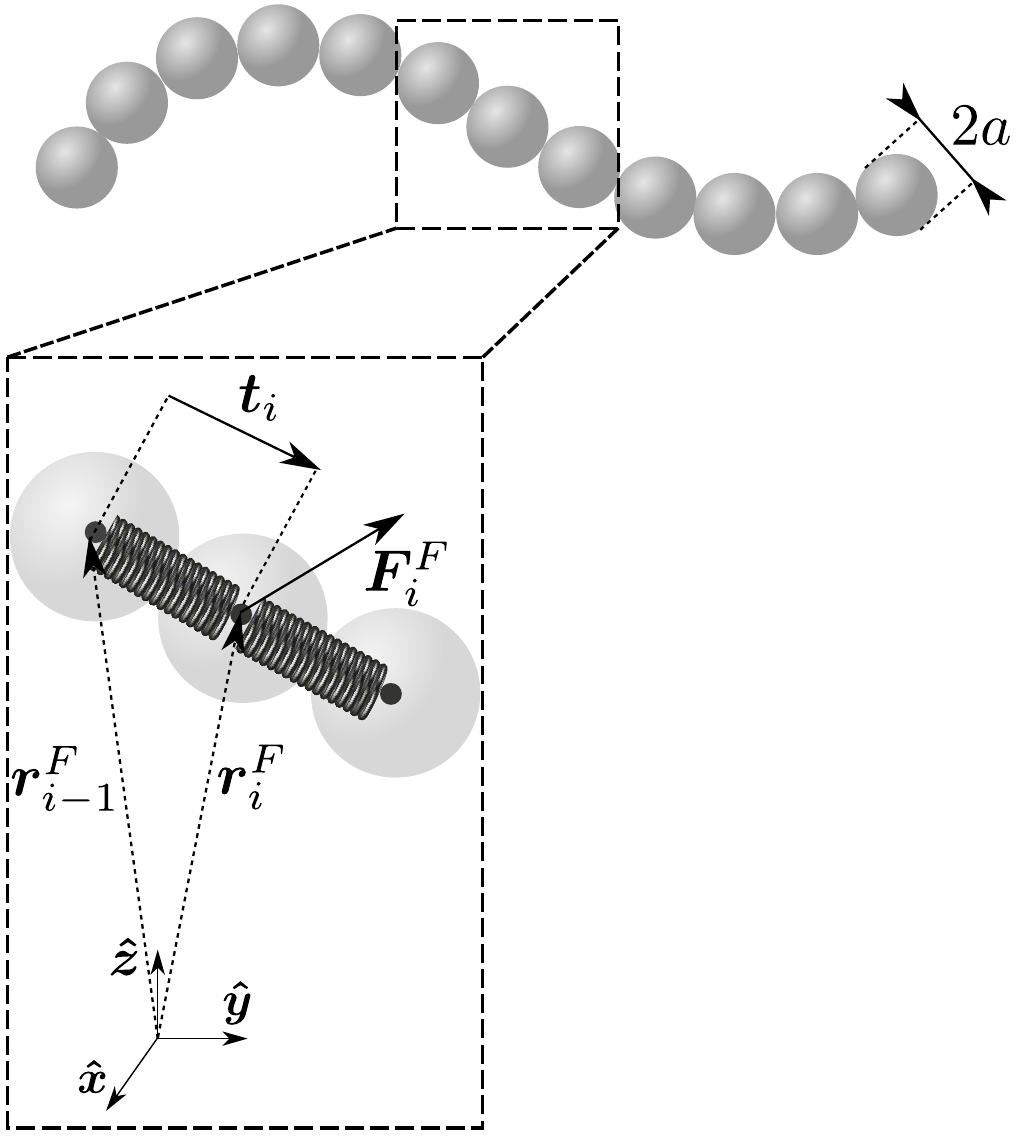}
    \caption{Discrete representation of a flexible fiber that is immersed in a viscous fluid. The fiber is made of evenly spaced $N_F$ markers located at $\bm{r}_i^{F}$ and connected by linear springs; see the inset. Each marker is subject to a total force $\bm{F}_i^F$ that includes the internal elastic forces and external forces, and $\bm{t}_i=\bm{r}_i^F - \bm{r}_{i-1}^F$ is the discrete tangent vector to the centerline. Furthermore, each marker is enclosed by a spherical bead of radius $a$, whose boundary corresponds to a region over which (i) the forces $\bm{F}_i^F$ are spread to the fluid, and (ii) the fluid velocity is interpolated.} 
    \label{fig: illustration_of_the_bead_spring_model}
\end{figure}

The fiber is discretized as a series of evenly spaced $N_F$ markers with position   $\bm{r}_i^{F} \in \mathbb{R}^3$ (see Fig.\ \ref{fig: illustration_of_the_bead_spring_model}) that are subject to internal elastic forces and external forces. Furthermore, two adjacent markers are connected by linkers made of springs, and the tangent vector to the centerline -- at the position of marker $i-1$ -- is defined such that $\bm{t}_i=\bm{r}_i^{F} - \bm{r}_{i-1}^{F}$(see Fig.\ \ref{fig: illustration_of_the_bead_spring_model}). Since we are interested in inextensible fibers, we prescribe the stretching moduli of the linkers (see Sec. \ref{sec: model_and_numerical_methods}) so as to allow only small variations $\Delta L$ in the fiber length $L$ -- less than $0.35\%$ of its initial length $L_0$.

The fiber being immersed in a viscous fluid, in the absence of inertia, the fluid velocity $\bm{u}(\bm{r})$ \footnote{The time dependence is omitted for notation convenience.} obeys the Stokes equations forced by regularized point forces $\bm{F}_j^F \in \mathbb{R}^{3}$, 

\begin{equation}
\left\{
\begin{aligned}
-\eta \nabla^2 \bm{u} + \bm{\nabla} p &=\sum_{j=1}^{N_F} \frac{1}{4\pi a_j^2}\bm{F}_j^F \left.\delta(|\bm{r} - \bm{r}_0| - a_j)\right|_{\bm{r}_0=\bm{r}_j^{F}}, \\
\bm{\nabla} \cdot \bm{u} &= 0,
\end{aligned} 
\right. 
\label{eq: forced_stokes_equations},
\end{equation}
where $\bm{r}_0$ is a \textit{source point}, $\bm{r}$ is a \textit{field point}, and the distribution $(4\pi a_j^2)^{-1}\delta(|\bm{r} - \bm{r}_0|-a_j)$ is the spherically-symmetric delta function or \textit{bead function} of width $\sim a_j$. In Eq.\ \eqref{eq: forced_stokes_equations}, $\bm{F}_j^F = \bm{F}_{G,j}^F + \bm{F}_{E,j}^F \in \mathbb{R}^{3}$ is the sum of the gravitational force, $\bm{F}_{G,j}^F$, and the elastic force, $\bm{F}_{E,j}^F$ (see Sec. \ref{sec: model_and_numerical_methods}). 

Owing to the linearity of the Stokes equations, the solution $\bm{u}(\bm{r})$ to Eq.\ \eqref{eq: forced_stokes_equations} is given as a superposition of convolutions -- over the boundary $S_j$ of the domain centered around the position of marker $j$  -- of the Green's function $\mathbb{G}(\bm{r}, \bm{r}_0) \in \mathbb{R}^{3 \times 3}$ with the regularized point forces, 

\begin{equation}
    \bm{u}(\bm{r}) = \sum_{j=1}^{N_F} \frac{1}{4\pi a_j^2}\int_{S_j}\mathbb{G}(\bm{r},\bm{r}_0)\cdot\bm{F}_j^F \delta(|\bm{r} - \bm{r}_0| - a_j) \mathrm{d}S(\bm{r}_0)\bigg|_{\bm{r}_0 =\bm{r}_j^F}.
\label{eq: fluid_velocity}
\end{equation}
On the other hand, the no-slip boundary condition gives the translational velocity of a given marker $\bm{U}_i \in \mathbb{R}^{3}$ as a convolution of the fluid velocity $\bm{u}(\bm{r})$ with the bead function, 
\begin{equation}
    \bm{U}_i^{F} = \frac{1}{4\pi a_i^2}\int_{S_i} \bm{u}(\bm{r}) \delta(|\bm{r} - \bm{r}_0| - a_i) \mathrm{d}S(\bm{r})\bigg|_{\bm{r}_0 =\bm{r}_i^F} . 
    \label{eq: linear_bead_velocity}
\end{equation}
Thus, by substituting Eq.\ \eqref{eq: fluid_velocity} into Eq.\ \eqref{eq: linear_bead_velocity}, we obtain the so-called \textit{mobility relation}, which gives the velocities of the markers in terms of the forces,

\begin{equation}
    \bm{U}_i^F = \sum_{j=1}^{N_F} \bm{\mathcal{M}}_{ij}\cdot\bm{F}_j^F,
    \label{eq: mobility_relation}
\end{equation}
where  $\bm{\mathcal{M}}_{ij} \in \mathbb{R}^{3 \times 3}$ is the Rotne-Prager-Yamakawa (RPY) tensor \cite{rotneVariationalTreatmentHydrodynamic1969,wajnrybGeneralizationRotnePragerYamakawaMobility2013, zukRotnePragerYamakawa2014},

\begin{equation}
\bm{\mathcal{M}}_{ij} = \frac{1}{4\pi a_i^2}\int_{S_i}\delta(|\bm{r} - \bm{r}_0| - a_i) \mathrm{d}S(\bm{r})\frac{1}{4\pi a_j^2}\int_{S_j}\mathbb{G}(\bm{r},\bm{r}_0)\delta(|\bm{r} - \bm{r}_0| - a_j) \mathrm{d}S(\bm{r}_0)\bigg|^{\bm{r}=\bm{r}_j^F}_{\bm{r}_0=\bm{r}_i^F} 
	 \label{eq: rpy}.
\end{equation}
The RPY tensor is symmetric positive definite (SPD), whether the boundaries $S_i$ and $S_j$ overlap or not, and can be evaluated in different domains: unbounded \cite{wajnrybGeneralizationRotnePragerYamakawaMobility2013, zukRotnePragerYamakawa2014}, half-space above a no-slip wall \cite{swanSimulationHydrodynamicallyInteracting2007a}, confined \cite{swanParticleMotionParallel2010}, and triply periodic \cite{mizerskiRotnePragerYamakawaApproximationPeriodic2014, fioreRapidSamplingStochastic2017}. In the latter, the suitable Green's function is given by Hasimoto \cite{hasimotoPeriodicFundamentalSolutions1959},  

\begin{equation}
\mathbb{G}(\bm{r}, \bm{r}_0) = \frac{1}{\eta V}\sum_{\bm{k} \neq \bm{0}}\frac{1}{\lVert\bm{k}\rVert^2}\big(\bm{I} - \hat{\bm{k}}\otimes\hat{\bm{k}}\big)e^{i\bm{k}\cdot (\bm{r} - \bm{r}_0)},
\label{eq: green_function_in_periodic_domain_bis}
\end{equation}
where $\bm{I} \in \mathbb{R}^{3\times 3}$ is the identity matrix, $V$ is the volume of the unit cell, and $\bm{k}$ are the wave vectors that are defined from the set of reciprocal vectors determining the unit cell, with $\hat{\bm{k}}=\bm{k}/\lVert\bm{k}\rVert$.

\subsection{Constrained formulation to account for the pillars: mixed mobility-resistance problem}

The fiber is settling through an array of rigid pillars. Each pillar is made of $N_O$ markers with position  $\bm{r}_i^{O} \in \mathbb{R}^3$, and we prescribe their motion to ensure they have zero degrees of freedom (DOF). This restriction leads to the following kinematic constraint equation:
\begin{equation}
     \bm{\mathcal{J}} \cdot \bm{U} = \bm{0},
    \label{eq: kinematic_constraint}
\end{equation}
where $\bm{U} = [\bm{U}_1^{F}, \ldots, \bm{U}_{N_F}^{F}, \bm{U}_{1}^{O}, \ldots,\bm{U}_{8N_O}^{O}]$ is a $3N=3(N_F+8N_O)$ vector collecting the translational velocities of the markers in the computational domain (see Sec. \ref{sec: model_and_numerical_methods}), and $\bm{\mathcal{J}} \in \mathbb{R}^{24N_O \times 3N}$ is the constraint Jacobian, which maps $\mathbb{R}^{3N}$ onto $\mathbb{R}^{24N_O}$. In other words, $\bm{\mathcal{J}}$ will select the $8N_O$ markers of the $8$ pillars in the computational domain to prescribe their velocities to zero. It can be constructed by blocks of $3 \times 3$ sub-matrices, such that 

\begin{equation}
\bm{\mathcal{J}}_{lk}=\left\{
\begin{aligned}
&\bm{I} \quad \text{for} \quad k=N_F + l\\
&\bm{0} \quad \text{otherwise}
\end{aligned} 
\right.  .
\label{eq: matrix_j}
\end{equation}
From classical mechanics, it is well known that linear nonholonomic constraints \eqref{eq: kinematic_constraint}, i.e., linear velocity-dependent constraints,  give rise to constraint forces $\bm{F}_{C} \in \mathbb{R}^{3N}$  of the form

\begin{equation}
    \bm{F}_{C} = \bm{\mathcal{J}}^{T} \cdot \bm{\lambda},
    \label{eq: constraint_forces}
\end{equation}
where  $\bm{\lambda} \in \mathbb{R}^{24N_O}$ is a set of Lagrange multipliers. Equation\ \eqref{eq: constraint_forces} stems from the principle of virtual work, since the constraint forces are workless \cite{lanczosVariationalPrinciplesMechanics1986}. 

The constraint forces \eqref{eq: constraint_forces} induce a disturbance motion, $\overline{\bm{U}} = \bm{\mathcal{M}}\cdot\bm{F}_{C}$, which corrects the unconstrained motion of the markers, $\widehat{\bm{U}}=\bm{\mathcal{M}}\cdot\bm{F}$, where $\bm{F} = \bm{F}_G + \bm{F}_E + \bm{F}_R  \in \mathbb{R}^{3N}$ (see Sec. \ref{sec: model_and_numerical_methods}) and $\bm{\mathcal{M}} \in \mathbb{R}^{3N \times 3N}$ \eqref{eq: rpy}. Thus, the solution to the coupling problem $\{\text{Fiber} \;+\; \text{Pillars}\}$ is therefore a superposition of the two aforementioned sub-problems,  

\begin{equation}
    \begin{aligned}
    \bm{U} &= \widehat{\bm{U}} + \overline{\bm{U}}\\
           &= \bm{\mathcal{M}} \cdot\bm{F} + \bm{\mathcal{M}} \cdot\bm{F}^{C} \\
            &= \bm{\mathcal{M}} \cdot \bm{F} + \bm{\mathcal{M}} \cdot \bm{\mathcal{J}}^{T} \cdot \bm{\lambda},
    \label{eq: mobility_in_the_system}
    \end{aligned}
\end{equation}
which leads to a \textit{mixed mobility-resistance} problem, as the Lagrange multipliers $\bm{\lambda}$ and the velocities $\bm{U}$ are both unknowns. Then, using the kinematic equation of motion \eqref{eq: mobility_in_the_system} and the kinematic constraint \eqref{eq: kinematic_constraint}, the coupling problem $\{\text{Fiber} \;+\; \text{Pillars}\}$  can be written in a compact matrix notation as a linear system to solve for $(\bm{\lambda}, \bm{U})$,

\begin{equation}
    \begin{bmatrix}
    -\bm{\mathcal{M}}\cdot\bm{\mathcal{J}}^{T} & \bm{I} \\
    \bm{0}                               & \bm{\mathcal{J}}\\ 
    \end{bmatrix}
    \cdot
    \begin{bmatrix}
    \bm{\lambda}\\
    \bm{U}
    \end{bmatrix}
    = 
    \begin{bmatrix}
    \bm{\mathcal{M}}\cdot\bm{F}\\
    \bm{0}
    \end{bmatrix}.
    \label{eq: mixed_mob_res_problem}
\end{equation}
It is straightforward to eliminate the Lagrange multipliers $\bm{\lambda}$ in the above system \eqref{eq: mixed_mob_res_problem}, in order to derive a constrained mobility problem, i.e., with the velocities $\bm{U}$ as the only unknowns,  
\begin{equation}
    \begin{aligned}
        \frac{\mathrm{d} \bm{R}}{\mathrm{d} t} &\equiv \bm{U}\\
        &= \left[\bm{I} - \bm{\mathcal{M}}\cdot\bm{\mathcal{J}}^{T}\cdot\bm{\mathcal{M}}_{OO}^{-1}\cdot\bm{\mathcal{J}}\right]\cdot\bm{\mathcal{M}}\cdot\bm{F}\\
        &= \bm{\mathcal{P}} \cdot \widehat{\bm{U}}\\
        &= \bm{\mathcal{N}} \cdot \bm{F},
        \label{eq: kinematic_equation_of_motion}
    \end{aligned}
\end{equation}
where  $\bm{R} = [\bm{r}_1^{F}, \ldots, \bm{r}_{N_F}^{F}, \bm{r}_{1}^{O}, \ldots,\bm{r}_{8N_O}^{O}]$. In the second line of Eq.\ \eqref{eq: kinematic_equation_of_motion}, $\bm{\mathcal{M}}_{OO} = \left[\bm{\mathcal{J}}\cdot\bm{\mathcal{M}}\cdot\bm{\mathcal{J}}^{T}\right] \in \mathbb{R}^{24N_O \times 24N_O}$ is the mobility matrix that encodes hydrodynamic interactions between the markers that are located on the pillars. In the third line, $\bm{\mathcal{P}}=\left[\bm{I} - \bm{\mathcal{M}}\cdot\bm{\mathcal{J}}^{T}\cdot\bm{\mathcal{M}}_{OO}^{-1}\cdot\bm{\mathcal{J}}\right] \in \mathbb{R}^{3N \times 3N}$ is a projector, i.e., $\bm{\mathcal{P}}^2=\bm{\mathcal{P}}$, which maps the unconstrained solution $\widehat{\bm{U}}\in \mathbb{R}^{3N}$ onto the constraint space $\mathbb{C} = \{\bm{U} \in \mathbb{R}^{3N} \mid \bm{\mathcal{J}} \cdot \bm{U} = \bm{0}\}$. Finally, the fourth line corresponds to the action of the constraint mobility matrix $\bm{\mathcal{N}}=\bm{\mathcal{P}} \cdot \bm{\mathcal{M}} \in \mathbb{R}^{3N \times 3N}$ on the applied forces $\bm{F}$.

At first sight, it might seem easier to solve the constrained mobility problem \eqref{eq: kinematic_equation_of_motion} as done in Ref.\ \cite{makangaObstacleinducedLateralDispersion2023}, instead of solving the mixed mobility-resistance problem \eqref{eq: mixed_mob_res_problem}, since one has to solve the former for $3N$ unknowns and the latter for $24N_O+3N$. In fact, the former is suitable for small systems, provided that the constraint mobility matrix $\bm{\mathcal{N}}$ can be formed efficiently. The main disadvantage is that numerical instabilities may be a concern when forming $\bm{\mathcal{N}}$, especially when the mobility matrix $\bm{\mathcal{M}}_{OO}$ is ill-conditioned. In contrast, we found that solving the mixed mobility-resistance problem \eqref{eq: mixed_mob_res_problem} iteratively, using a preconditioned GMRES method, is numerically stable and suitable for large systems of fibers and pillars. As a matter of fact, such an approach has a computational complexity that is quadratic in the number of markers $N$ compared to direct methods, which scale as $\mathcal{O}(N^3)$. This efficiency is mainly due to the cost of each iteration and the modest number of iterations required to solve \eqref{eq: mixed_mob_res_problem} when an appropriate preconditioner is used \cite{makangaTransportDeformationFlexible2023}.

\subsection{Algorithm and implementation}

To sum up, our numerical algorithm is as follows:

\begin{itemize}
    \item Initialize the positions of the markers  $\bm{R}$. 
    \item For each time step in the time loop:
    \begin{enumerate}
        \item Solve the mixed mobility-resistance problem \eqref{eq: mixed_mob_res_problem} iteratively, using a preconditioned GMRES method, for $(\bm{\lambda}, \bm{U})$.
        \item Update the positions of the markers $\bm{R}$ by integrating in time the kinematic equation of motion \eqref{eq: kinematic_equation_of_motion} (first line). Here, we use an implicit scheme based on a backward differentiation formula (BDF) with adaptive time stepping \cite{brownVODEVariableCoefficientODE1989}.  
    \end{enumerate}
\end{itemize}

The cost of each GMRES iteration while solving the mixed mobility-resistance problem \eqref{eq: mixed_mob_res_problem} depends on the action $\bm{\mathcal{M}}\cdot (\bullet)$ of the mobility matrix $\bm{\mathcal{M}}$ on a given vector. An efficient algorithm is therefore needed to compute the latter. Here, for a triply periodic domain, we use a GPU implementation of the Positively-Split-Ewald (PSE) method, whose detailed description can be found elsewhere \cite{fioreRapidSamplingStochastic2017, perezpelaezComplexFluidsGpu2022, fishLibMobilityPythonLibrary2026}.

\begin{figure}[h]
   \centering
    \includegraphics[width=0.55\textwidth]{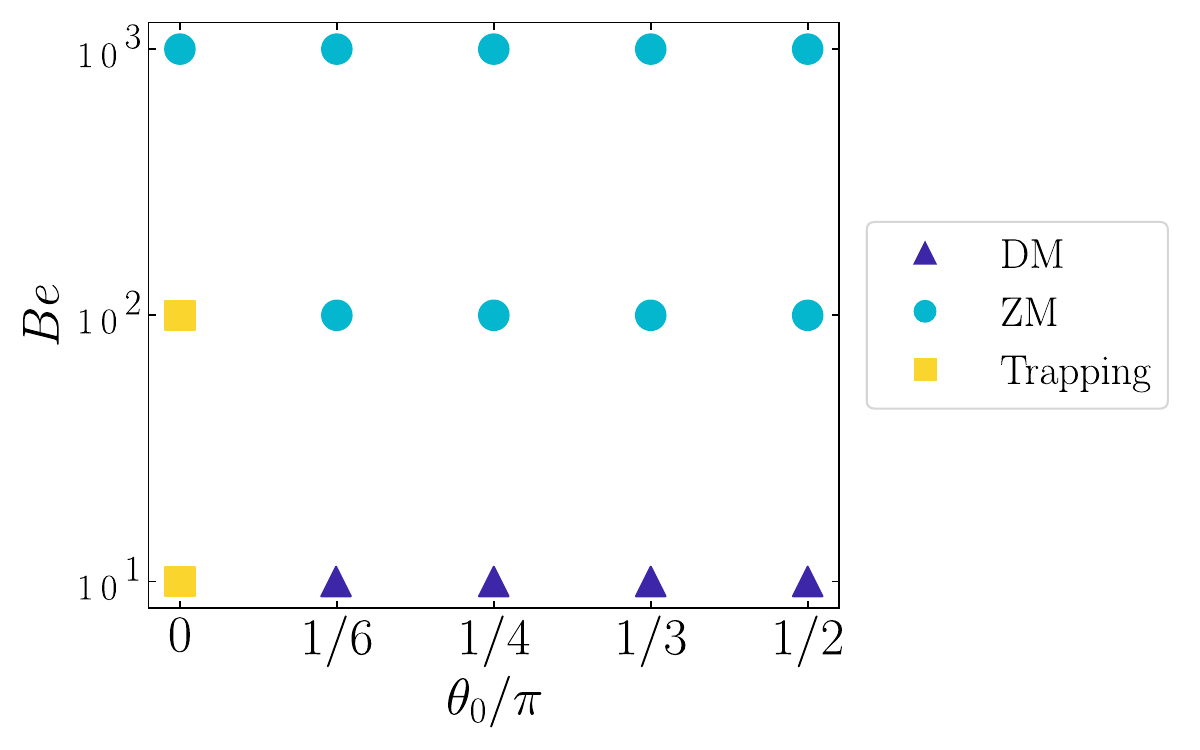}
    \caption{Phase diagram showing the migration modes for various values of the initial orientation, $\theta_0$ $(0 \leq \theta_0 \leq \pi/2)$, various values of the elastogravitational number, $Be$ $(10\leq Be \leq 1000)$, and a fixed value of the relative fiber length, $\xi = 2.36$. DM and ZM stand for ``Displacement Mode" and ``Zigzag Mode", respectively.} 
    \label{fig: phasePlotVariousBeAndThetaInitial}
\end{figure}

\section{Effects of the initial orientation of the fiber on the migration modes}
\label{app: effects_initial_orientation_migration}

To verify the robustness of our findings with respect to the initial orientation of the fiber, $\theta_0$, we perform long-time numerical simulations with an initially straight filament for various values of $\theta_0$, various values of the elastogravitational number, $Be$, and a fixed value of the relative fiber length, $\xi$. Figure \ref{fig: phasePlotVariousBeAndThetaInitial} shows the different migration modes for the explored parameter space, $0 \leq \theta_0 \leq \pi/2$, $10\leq Be \leq 1000$, and $\xi=2.36$. We find that the migration modes do not depend on the initial orientation, provided that the fiber is not initially aligned with gravity, i.e., $\theta_0 = 0$; in the latter case, the fiber remains trapped in the array at short times for $Be=10$ and $Be=100$ (see Movies 8 and 9 in the Supplemental Material \cite{SSM}). 

\bibliography{draft}

\end{document}